\documentclass[aps,prl,reprint]{revtex4-2}

\usepackage{graphicx}
\usepackage{dcolumn}
\usepackage{bm}
\usepackage{hyperref}
\usepackage[mathlines]{lineno}
\usepackage{amssymb, amsmath, xcolor, pifont, soul}
\usepackage{color}
\usepackage{hyperref}
\usepackage{booktabs}
\usepackage{tabu}
\usepackage{makecell}
\usepackage{booktabs}

\begin{document}


\title{Al-Based Few-Hydrogen Metal-Bonded Perovskite High-$T_c$ Superconductor  Al$_4$H up to 54 K under Atmospheric Pressure}

\author{Yong He}
\author{Jing Lu}
\author{Xinqiang Wang}
\author{Jun-jie Shi} \email{Corresponding author. \\ jjshi@pku.edu.cn}
\affiliation{State Key Laboratory for Artificial Microstructures and Mesoscopic Physics, School of Physics, Peking University Yangtze Delta Institute of Optoelectronics, Peking University, Beijing 100871, China}

\date{\today}

\begin{abstract}
Multi-hydrogen lanthanum hydrides have shown the highest critical temperature $T_c$ at 250-260 K under 170-200 GPa. However, such high pressure is a great challenge for sample preparation and practical application. To address this challenge, we propose a novel design strategy for ambient-pressure high-$T_c$ superconductors by constructing new few-hydrogen metal-bonded perovskite hydrides, such as Al-based superconductor $\rm {Al_4H}$, with better ductility than the well-known multi-hydrogen, cuprate and Fe-based superconductors. Based on the Migdal-Eliashberg theory, we predict that the structurally stable $\rm {Al_4H}$ has a favorable high $T_c$ up to 54 K under atmospheric pressure, similar to SmOFeAs.
\end{abstract}

\maketitle


Hydrogen, the lightest element with the smallest radius, has been predicted to possess the highest superconducting transition temperature $T_c$$\sim$356 K with a metallic modification at an ultrahigh pressure near 500 GPa~\cite{Ashcroft1968, Jeffrey2011}. Unfortunately, the experimental observation on metallic hydrogen has not yet been confirmed even at about 400 GPa~\cite{Monserrat2018, Loubeyre2020}. In 2004, Ashcroft suggested that the hydrogen dominant material, i.e., multi-hydrogen hydride, could be a powerful candidate for high-$T_c$ superconductors under moderate pressure where the combination of hydrogen and other elements can produce chemical precompression to reduce the pressure of hydrogen metallization~\cite{Ashcroft2004}. Following this idea, many multi-hydrogen covalent hydrides have been predicted by using structural search methods at high pressure and some of them have been synthesized by using diamond anvil cell~\cite{Flores-Livas2020}. Duan \textit{et al.} predicted that $\rm {(H_2S)_2H_2}$ with $Im$-$3m$ structure possesses high $T_c$ of 191-204 K under 200 GPa~\cite{Duan2014}, confirmed by experiments~\cite{Drozdov2015}. Several multi-hydrogen materials, such as SiH$_4$~\cite{Eremets2008}, YH$_{9}$~\cite{Kong2021}, LaH$_{10}$~\cite{Drozdov2019, Somayazulu2019}, ThH$_{10}$~\cite{Semenok2020} and $\rm {CaH_6}$~\cite{Ma2022}, have been synthesized successively, and their superconductivity with $T_c$ about 17, 243, 250-260, 161 and 215 K has been certified under high pressure. The superconductivity of Th$_4$H$_{15}$ and (ThM$_x$)$_4$H$_{15}$ (M=Ce, La, Zr, Bi and Y) was observed at $T_c$$\sim$8.05-8.35 and $T_c$$\leq$7.5 K at ambient pressure~\cite{Satterthwaite1970} and 50 atm~\cite{Oesterreicher1977}, respectively. Based on first-principles simulations, Vocaturo \textit{et al.} predicted that $\rm PdCuH_2$ is a superconductor with $T_c$$\sim$34 K at ambient pressure~\cite{Vocaturo2022}. The estimated $T_c$ in CrH (CrH$_3$) is 10.6 ($\sim$37.1) K at atmospheric pressure (81 GPa)~\cite{Yu2015}. As a high-temperature superconductor, KB$_2$H$_8$ is predicted with $T_c$$\sim$134-146 K around 12 GPa~\cite{Gao2021}.

\begin{figure*}[t]
\includegraphics[width=14cm]{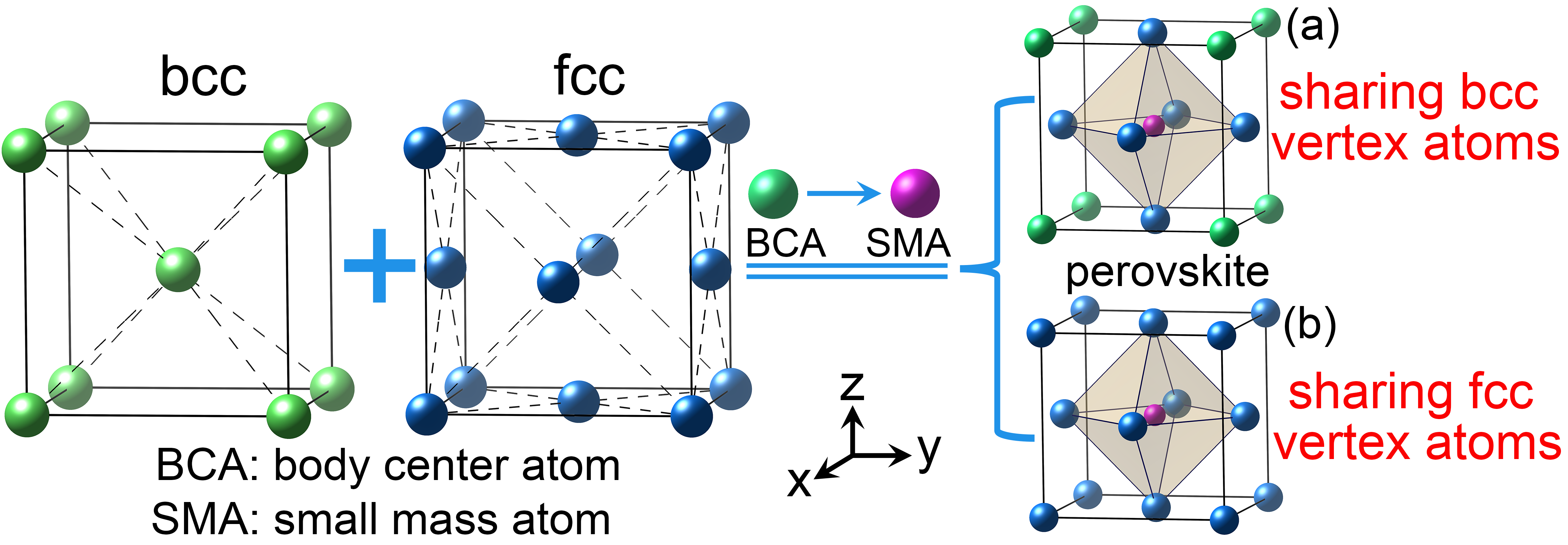}
\caption{\label{Fig_1}The design idea of few-hydrogen metal-bonded perovskites, formed by combining bcc metal lattice (left panel) with the body-centered atom replaced by small mass or radius atom and fcc metal lattice (central panel), with a much larger packing fraction than fcc lattice. The right panel (a) depicts ternary perovskite (see Table~S3 for several examples), derived from bcc and fcc lattices by sharing bcc vertex atoms. If sharing fcc vertex atoms, binary perovskite (right panel (b)) can be obtained. The fcc body center, i.e., octahedral interstice O$_{\rm h}$, can be regarded as a normal lattice point from the viewpoint of perovskite. If the body center of a fcc metal, such as aluminum, is occupied by H atom with the lightest mass and smallest radius, a typical binary Al-based metal-bonded perovskite Al$_4$H can thus be formed naturally.}
\end{figure*}

Among multi-hydrogen materials, aluminum hydrides have been recognized as important hydrogen storage and superconducting materials. As one of the most promising hydrogen storage materials, $\rm {AlH_3}$ has been extensively studied due to its high hydrogen storage and rapid dehydrogenation capacity~\cite{Yu2021}. The calculated $T_c$ in $\rm {AlH_3}$, a poor superconductor due to the absence of conduction electrons caused by Al$^{+3}$ and H$^{-1}$, rapidly decreases with compression and becomes less than 2.04 K at 110 GPa~\cite{Wei2013, Islam2010} and eventually approaches zero at above 210 GPa~\cite{Abe2019}, confirmed by experiments~\cite{Goncharenko2008}. The superconductivity of $\rm {AlH_2}$ with $T_c$=6.75 K, dominated by the electron-optical phonon coupling~\cite{Wei1987}, has been observed in experiments~\cite{Lamoise1975}. The calculated $T_c$ of AlH with $R\overline{3}m$ symmetry is about 58 K under 180 GPa, and the estimated $T_c$ of $\rm {Al_2H}$ with $P\overline{3}m$1 structure is 3.5 K at 195 GPa~\cite{Abe2019}. Obviously, the superconductivity of few-hydrogen aluminum hydrides is still absent under ambient pressure.

Considering that both the face-centered cubic (fcc) and body-centered cubic (bcc) lattices are very common metal crystal structures and throughly analyzing the relationship (see Eq.~\ref{equ_1}) among fcc, bcc and perovskite structures,
\begin{equation}\label{equ_1}
 \rm {fcc+bcc\xrightarrow{vertex-sharing}perovskite},
\end{equation}
we propose a novel idea by constructing a new few-hydrogen metal-bonded perovskite hydride derived from fcc metal with the smallest radius H atom at its body center as an ambient-pressure high-$T_c$ superconductor, different from the semiconducting and metallic perovskites featured with ionic and covalent bonds, which can be regarded as a complete reversal of the conventional design strategy of multi-hydrogen high-$T_c$ superconductors at several hundred GPa, which is impractical. Because aluminum is the only material with good ductility, low density, ultrahigh abundance and environmental friendliness for long-distance power transmission, it would bring great benefits to humanity if becoming a high-$T_c$  superconductor. As an application of our new idea, binary Al-based metal-bonded perovskite $\rm {Al_4H}$ is chosen, in which the body-centered H atom, occupied the “sphere interstice” of Al atom lattice, is naturally fixed by the stable fcc skeleton of Al atoms without any extrinsic pressure. The electronic band structure, phonon dispersion, electron-phonon coupling (EPC) and superconducting property have been investigated theoretically. Excellent superconducting characteristics and ductility of $\rm {Al_4H}$ are confirmed at ambient pressure, which is favorable for motivating experimental investigations and even practical applications in the future.

Because many metals adopt fcc and bcc structures, we thus theoretically design few-hydrogen metal-bonded hydrides to realize high-$T_c$ superconductors at ambient pressure. Our novel idea is illustrated in Fig.~\ref{Fig_1}, where the left panel shows the bcc conventional cell with six intrinsic vacancies at its face centers, and the central panel indicates the fcc conventional cell with a natural vacancy at its body center. Comparing fcc with bcc structures, the body center of fcc structure can be easily occupied by small radius atom, such as H, He or B atom, to form metal-bonded perovskite structure (right panel of Fig.~\ref{Fig_1}). Naturally, we choose H atom to occupy the body center of fcc aluminum, an environment-friendly and low-cost superconductor with $T_c$$\sim$1.20 K~\cite{Cochran1958}, to form $\rm {Al_4H}$ perovskite (see Table~S1 and Fig.~S4 of Supplemental Material~\cite{SupplementalMaterial} for its lattice constant, Wyckoff position, Bader charge and XRD pattern~\cite{Momma2011}). Obviously, two key issues, closely related to multi-hydrogen high-$T_c$ superconductors, i.e., ultrahigh pressure and poor ductility caused by the covalent bond, can be solved by our designed new few-hydrogen metal-bonded perovskites under atmospheric pressure.

\begin{figure}[ht]
\includegraphics[width=8.4cm]{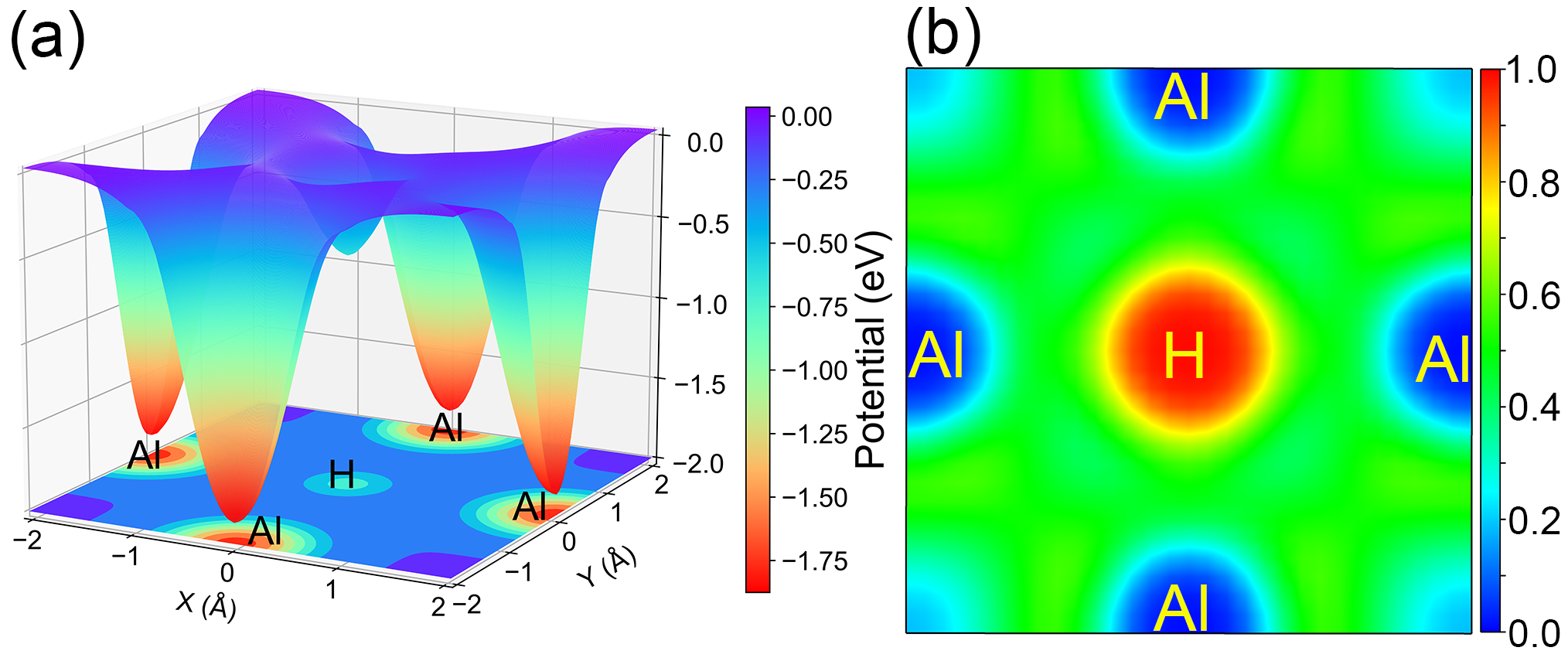}
\caption{\label{Fig_2}Calculated electrostatic potential (a) and electron localization function (b) on the Al-H plane.}
\end{figure}

Considering that the structural stability of perovskite Al$_4$H is the foundation of its superconductivity, we thus comprehensively check its stability from crystal structure, formation energy, $ab$ $initio$ molecule dynamics simulation, phonon spectra and previous experiments for hydrogen implantation in fcc aluminum with two highly competitive interstitial sites, i.e., octahedral ($\rm O_h$) and tetrahedral ($\rm T_d$)~\cite{Kresse1996, Blochl1994, Perdew2008, Nose1991, Wei1987_2, Bugeat1976, Bugeat1979, Schluter1993, McLellan1983, Papp1977, Lamoise1975, Hashimoto1983, Ambat1996, Mills1965, Drozdov2019, Jmerik2009, Perdew1996, Wolverton2004, Armiento2005} (see sections II and III, Figs. S1, S2, S5-S8 and Table S2 of Supplemental Material~\cite{SupplementalMaterial}). Figure~\ref{Fig_2} (a) shows the electrostatic potential on the Al-H plane in Al$_4$H, indicating that the H atom stays stably at the bottom of its potential well. After optimizing the crystal structure and calculating the zero-point energy by using several advanced functionals including PBEsol~\cite{Perdew2008} and AM05~\cite{Armiento2005}, we find that the body center $\rm O_h$ has the lowest energy, clarifying the long-running debate on the priority of $\rm O_h$ and $\rm T_d$ sites for H atom occupation in aluminum~\cite{Popovic1974, McLellan1983, Bugeat1979, Wolverton2004}. The stability of perovskite Al$_4$H was further confirmed by using a $disorder$ code with enumerated method (Fig. S5)~\cite{Lian2020}. We believe from our careful calculations that few-hydrogen binary metal-bonded perovskite Al$_4$H with H atom at $\rm O_h$ site is the most stable structure under the preconditions of fcc aluminum and Al:H=4:1.

To clarify the nature of chemical bonds in Al$_4$H, we further investigate its electron localization function (Fig.~\ref{Fig_2} (b)), i.e., the normalized electron density, which intuitively shows the bond characteristics between Al and H atoms. We find that both Al and H ions are immersed in an almost uniform electron sea with the normalized electron density of 0.6, indicating a typical metallic bond combination, different from multi-hydrogen covalent superconductors at high pressure. Obviously, these conduction electrons are mainly derived from 3$s$ and 3$p$ valence electrons of Al atom. It is the Al-H metallic bond that enhances structural stability of Al$_4$H and guarantees its better ductility than the multi-hydrogen, cuprate and Fe-based superconductors.

\begin{figure}[h]
\includegraphics[width=8.4cm]{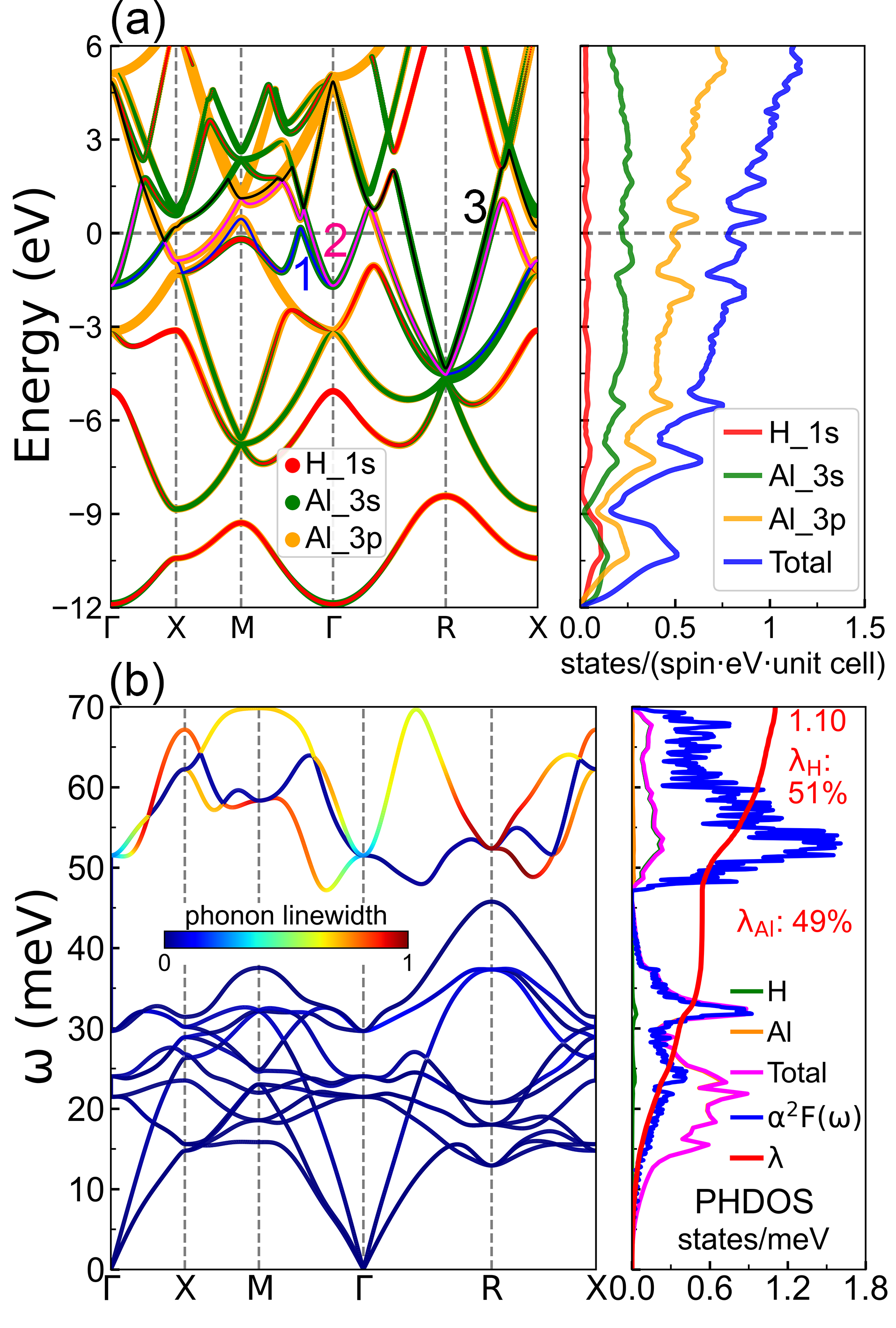}
\caption{\label{Fig_3}Energy band and corresponding projected DOS (a) and lattice dynamic property (b) of Al$_4$H. The size of the colored circles shows the orbital character, which is proportional to the weight for the orbital contribution. (b) Left: Phonon spectra decorated with the normalized phonon linewidth. Right: The projected phonon DOS (PHDOS), Eliashberg spectral function ${\alpha}^2F({\omega})$ and EPC constant ${\lambda}$=${\lambda}_{\rm Al}$+${\lambda}_{\rm H}$.}
\end{figure}

\begin{figure}[ht]
\includegraphics[width=8.4cm]{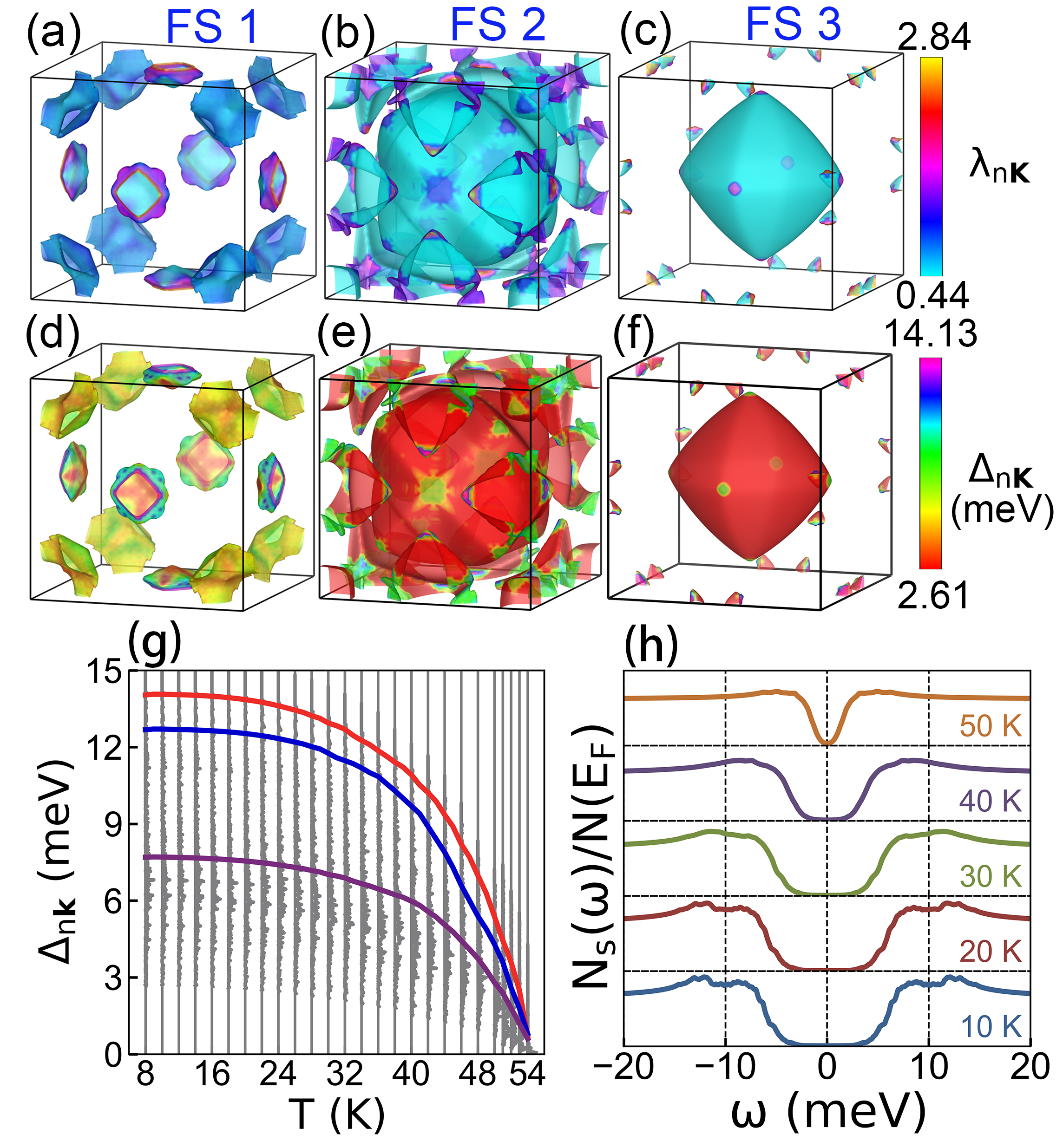}
\caption{\label{Fig_4}EPC constant and superconducting gap of Al$_4$H. (a)-(c) Fully $\textbf{k}$-resolved EPC strength $\rm {\lambda}_{n\mathbf{k}}$ on the FS for three crossing bands~1, 2 and 3 (Fig.~\ref{Fig_3} (a)). (d)-(f) Fully $\textbf{k}$-resolved superconducting gap $\rm {{\Delta}_{n\mathbf{k}}}$ on the three FS sheets at 10 K. (g) The anisotropic superconducting gap $\rm {{\Delta}_{n\mathbf{k}}}$ as a function of temperature, where the red, blue and purple lines represent the gaps defined by the maximum normalized quasiparticle DOS. (h) The normalized quasiparticle DOS from 10 to 50 K.}
\end{figure}

Let us now investigate electronic structures of Al$_4$H. Figure~\ref{Fig_3} (a) shows our calculated electronic bands with the resolved atomic orbital contribution~\cite{Giannozzi2009, Giannozzi2017, Giannozzi2020, Perdew2008, Martin2015}, indicating a typical metal energy band with three bands, labeled as 1, 2 and 3, crossing the Fermi level ($E_{\rm F}$=0 eV). Figure~S2 (a) gives the corresponding Wannier interpolation bands.  Band~1 is mainly dominated by Al 3$s$ and 3$p$ orbitals, consistent with Fig.~S10 (d) and (g), which has three camel shapes crossing $E \rm {_F}$ along the X-M-$\Gamma$, M-$\Gamma$ and $\Gamma$-R directions, respectively. Band 2 crosses $E \rm {_F}$ along the whole high symmetry path. These multiple crossings at $E \rm {_F}$ give birth to a complex Fermi surface (FS) sheet and the orbital characteristics can be derived from H 1$s$, Al 3$s$ and 3$p$ states, in good agreement with Fig.~S10 (b), (e) and (h). Band 3, dominated by Al 3$s$ and 3$p$ orbitals (see Fig.~S10 (f) and (i)), is also shown near the Fermi level. We can deduce from the projected density of states (DOS) of Fig.~\ref{Fig_3} (a) (right panel) that the 3$s$ and 3$p$ orbitals of Al atom contribute almost 31.7\% and 64.5\% to the total DOS, while the contribution from the H 1$s$ orbital is about 3.8\%. It is three FS sheets induced by three bands crossing $E \rm {_F}$ that dominate three distinct superconducting gaps in Al$_4$H (Fig.~\ref{Fig_4} (d)-(g))~\cite{Floris2007}.

Before investigating superconductivity, we carefully assess the anharmonic effects of Al$_4$H in section IV of Supplemental Material~\cite{SupplementalMaterial} and find from Figs.~S8 and~S9 that Al$_4$H only exhibits weak anharmonicity~\cite{Errea2013, Errea2015, Errea2020, Hou2021, Yin1980, Rush1984}. We thus calculate the harmonic phonon spectra and EPC parameters of Al$_4$H by using density-functional perturbation theory~\cite{Baroni2001} and the Wannier interpolation technology on very fine $\textbf{q}$ and $\textbf{k}$ grids, as implemented in the EPW code~\cite{Giustino2007, Ponce2016, Giustino2017, Pizzi2020}. Figure~\ref{Fig_3} (b) shows phonon spectra embellished with the phonon linewidth, PHDOS, Eliashberg spectral function ${\alpha}^2F(\omega)$ and EPC strength ${\lambda}$ determined by the phonon linewidth, closely related to the electron-phonon interaction matrix element describing the scattering probability amplitude of an electron on FS by a phonon with wave vector $\textbf{q}$~\cite{Gao2015}. The calculated phonon spectra by using several different pseudopotentials including ONCV~\cite{Martin2015}, USPP and PAW~\cite{Andrea2014} are comprehensively compared with the all-electron spectra~\cite{elk} (Fig.~S1). The commonly-used acoustic sum rules, such as ``simple" and ``crystal", are also carefully checked. We can find from Figs.~\ref{Fig_3} (b) and~S2 (b) that phonon spectra are divided into two parts separated by a small phonon energy gap. The high (low) frequency modes are dominated by H (Al) atom vibration. Compared with the maximum frequency of aluminum~\cite{Giaremis2021}, the cutoff frequency of Al$_4$H has an increase of 60\% due to vibration of H atom with the lightest mass $m_{\rm H}$$\sim$$m_{\rm Al}/27$, which is highly desirable to improve $T_c$. The resolved phonon linewidth of all branches below 46 meV indicates that they have a similar EPC strength, except the mode around 31 meV with a slightly larger phonon linewidth. Compared with the low-frequency acoustic modes, the H-related high-frequency optical phonons have a much stronger EPC because of their large phonon linewidths.

We further investigate the Eliashberg spectral function ${\alpha}^2F({\omega})$ using the maximally localized Wannier function interpolation technology and accumulate the EPC constant ${\lambda}$ (right panel of Fig.~\ref{Fig_3} (b)). The ${\alpha}^2F({\omega})$ exhibits a peak centered at 31 meV and a much larger main peak around 52 meV, which reveals great EPC due to the large phonon linewidth. The cumulative ${\lambda}={\lambda}_{\rm Al}+{\lambda}_{\rm H}$ is 1.10, in which contributions from Al and H atoms are 49\% and 51\%, respectively. We find that $\lambda$ in Al$_4$H is almost three times larger than that of aluminum (0.36)~\cite{Giaremis2021}, and superior to $\rm {MgB_2}$ (0.75)~\cite{Margine2013}, because of the large phonon linewidth of the high-frequency H-vibration modes and slight softening of the low-frequency Al-vibration modes.

It is worthwhile to note that, as the source of all novel properties of Al-based superconductor Al$_4$H, the H atom has a larger ${\lambda}_{\rm H}$ than ${\lambda}_{\rm Al}$, approaching 51\% of the total EPC constant (Fig.~\ref{Fig_3} (b)). The key role of H atom in the superconductivity of Al$_4$H can be analyzed as follows. Although the body-centered H atom, with a lower energy level than Fermi level (Fig.~\ref{Fig_3} (a)) and a much larger electronegativity than that of Al atom, takes electrons from its neighboring Al atoms to be localized around it (Fig.~\ref{Fig_2} (b) and Table~S1), the superconductivity of Al$_4$H is slightly affected by this electron transfer finally because the total DOS on FS decreases very little from 0.82 in Al to 0.78 states/(spin$\cdot$eV$\cdot$unit cell) in Al$_4$H. On the contrary, H atom in Al$_4$H induces new high-frequency optical phonons and greatly modifies the phonon spectra of aluminum (Fig.~\ref{Fig_3} (b)). Most importantly, these new H-related high-frequency optical phonons have the largest phonon linewidth among all phonons, directly leading to a larger EPC constant ${\lambda}_{\rm H}$ than ${\lambda}_{\rm Al}$ due to strong EPC between the H-related phonon and conduction electron~\cite{Nedrud1981, Yasutami1982, Yasutami2009}. This is extremely important to enhance $T_c$ of Al$_4$H indeed~\cite{Allen1975}. It is the large ${\lambda}_{\rm H}$ together with ${\lambda}_{\rm Al}$ that dominate the high $T_c$ in Al$_4$H.

To evaluate the anisotropy of EPC strength in Al$_4$H, we explored the $\textbf{k}$-resolved EPC constant ${\lambda}_{n\mathbf{k}}$ defined by $\lambda_{n \mathbf{k}}=\sum_{m \mathbf{k}^{\prime}, \nu} \frac{1}{\omega_{\mathbf{q} v}} \delta\left(\epsilon_{m \mathbf{k}^{\prime}}\right)\left|g_{n \mathbf{k}, m \mathbf{k}^{\prime}}^{\nu}\right|^{2}$~\cite{Margine2013}. Figure~\ref{Fig_4} (a)-(c) show the variation of ${\lambda}_{n\mathbf{k}}$ on the three FS sheets. The values of ${\lambda}_{n\mathbf{k}}$ on FS 1, 2 and 3 have a wide range of 0.45-2.84, 0.45-2.44 and 0.44-1.00, respectively, indicating the strong anisotropy inside each single sheet. Figure~S11 also plots the normalized distribution of ${\lambda}_{n\mathbf{k}}$ with a wide range of 0.45-2.84, further revealing strong anisotropy. We find that ${\lambda}_{n\mathbf{k}}$ on FS~1 is larger than on FS~2 and~3. Particularly, six flakes at the face centers of FS~1 exhibit the largest EPC strength because of Al 3$s$ and 3$p$ electrons strong coupling with phonons (Fig.~S10). The EPC constant ${\lambda}_{n\mathbf{k}}$ mainly comes from H 1$s$ and Al 3$p$ orbitals. Whereas, FS~3, dominated by the Al 3$s$ state, has a small contribution to ${\lambda}_{n\mathbf{k}}$ among three FS sheets.

Furthermore, the superconducting gap ${{\Delta}_{n\mathbf{k}}}$ of Al$_4$H has also been calculated  on the FS by numerically solving the anisotropic Migdal-Eliashberg equations with ${\mu}^*$=0.1~\cite{Carbotte1990, Allen1983, Margine2013}. In Fig.~\ref{Fig_4} (d)-(f), we show the $\textbf{k}$-resolved superconducting gap at 10 K on the three FS sheets. The gaps on FS~1, 2 and~3 are in the range of 2.61-14.13, 2.67-13.90 and 2.64-7.70 meV, respectively, indicating the strong anisotropy of ${{\Delta}_{n\mathbf{k}}}$ on each single FS sheet with a large vertical energy spread (Figs.~\ref{Fig_4} (g) and~S3). The larger the superconducting gap ${{\Delta}_{n\mathbf{k}}}$, the higher the critical temperature $T_c$. It is interesting to note that the distribution of ${{\Delta}_{n\mathbf{k}}}$ is very similar to that of ${{\lambda}_{n\mathbf{k}}}$,  suggesting that Al$_4$H is a phonon-mediated high-$T_c$ superconductor.

Figure~\ref{Fig_4} (g) presents the superconducting gap ${{\Delta}_{n\mathbf{k}}}$ as a function of temperature, in which three fully anisotropic superconducting gaps originating from the three FS sheets are featured with a remarkable vertical energy spread, different from aluminum with a single gap~\cite{Biondi1959, Douglass1961}. The red, blue and purple lines show the gaps corresponding to the maximum superconducting DOS on FS~1, 2 and~3 (Fig.~\ref{Fig_4} (h)), respectively. The three superconducting gaps belong to different FS sheets and always overlap with each other due to their large vertical energy spread and strong anisotropic EPC (Figs.~\ref{Fig_4} (a)-(c) and~S10). As temperature increases, the superconducting gaps decrease and eventually disappear at $T_c$=54 K (51 K with ${\mu}^*$=0.13, Fig.~S12), which is approximately 45 times larger than aluminum (1.20 K)~\cite{Cochran1958}, and higher than $\rm {Nb_3Ge}$ (23 K)~\cite{Testardi1974} and $\rm MgB_2$ (39 K)~\cite{Nagamatsu2001}, similar to Fe-based superconductor SmOFeAs (55 K)~\cite{Hideo2018}. We further obtain $T_c$=42 K from the isotropic Migdal-Eliashberg equations. Obviously, both the anisotropic effect and multiband gaps, closely related to the strong anisotropy of EPC, greatly enlarge $T_c$ about 28.6\%, which shows that the anisotropic effect is very important to improve $T_c$, as confirmed by Ref.~\cite{Floris2007}. We also find that the average value of three anisotropic single gaps is close to the isotropic single gap at T$\rightarrow$0 K.

Moreover, we calculate the normalized quasiparticle DOS $N_S({\omega})$ at different temperatures based on the formula $\frac{N_{S}(\omega)}{N({E\rm_F})}=\operatorname{Re}[\frac{\omega}{\sqrt{\omega^{2}-\Delta^{2}(\omega)}}]$~\cite{Margine2013}, where $N({E\rm_F})$ represents DOS of the normal state at Fermi level. We can see from Fig.~\ref{Fig_4} (h) that the superconducting gap decreases if temperature increases. The normalized quasiparticle DOS, as a function of excitation energy, exhibits three peaks, closely related to three different energy gaps in Al$_4$H. Furthermore, to prove that our novel idea (Fig.~\ref{Fig_1}) is fruitful, we find that, besides Al$_4$H, several other binary and ternary few-hydrogen metal-bonded perovskite M$_4$H (M=Ca, Cu, Rh) and AHM$_3$ (A=V, Nb, M=Al, Rh, Ca) hydrides are also phonon-mediated superconductors under ambient pressure (Table S3). Finally, the specific heat and critical magnetic field are investigated (Fig. S13)~\cite{John1964, Carbotte1990, Choi2003, Margine2013, Cochran1958, Corsan1969}.

In summary, we proposed a novel idea and found a new class of materials named as few-hydrogen metal-bonded perovskite hydrides, such as Al-based superconductor Al$_4$H, as strong anisotropic phonon-mediated high-$T_c$ superconductors at ambient pressure, by combining first-principles calculations and Wannier interpolation method, and solving the Migdal-Eliashberg equations. The structural stability of Al$_4$H is confirmed. As an environmental friendly and low-cost material, Al$_4$H, featured with metallic bonds, is the only new high-$T_c$ superconductor with both better ductility than multi-hydrogen, cuprate and Fe-based superconductors combined by covalent and ionic bonds and high-$T_c$ up to 54 K, similar to SmOFeAs. Therefore, contrary to the conventional design method of multi-hydrogen covalent high-$T_c$ superconductors at high pressure, our novel idea paves a new way for designing few-hydrogen metal-bonded hydride high-$T_c$ superconductors with simple structure under atmospheric pressure. We hope that this Letter will rekindle enthusiasm for predicting new materials with favorable superconducting property among hydrides at ambient pressure and can stimulate further experimental investigation in the near future.

\begin{acknowledgments}
This work was supported by the Beijing Outstanding Young Scientist Program (BJJWZYJH0120191000103) and the National Natural Science Foundation of China (61734001 and 61521004). We used the High Performance Computing Platform of the Center for Life Science of Peking University.
\end{acknowledgments}


%


\pagebreak
\begin{widetext}
\setcounter{equation}{0}
\setcounter{figure}{0}
\setcounter{table}{0}
\setcounter{page}{1}
\makeatletter
\renewcommand{\thefigure}{S\arabic{figure}}
\renewcommand{\theequation}{S\arabic{equation}}
\renewcommand{\thetable}{S\arabic{table}}

\begin{center}
\textbf{\large Supplemental Material for “Al-Based Few-Hydrogen Metal-Bonded Perovskite High-$T_c$ Superconductor Al$_4$H up to 54 K under Atmospheric Pressure”}
\end{center}

The present supplemental material provides more details about the theoretical and computational methods, structural stability and superconducting property in Al-based few-hydrogen binary metal-bonded perovskite hydride $\rm {Al_4H}$ at ambient pressure, including six sections, i.e., I. Computational methods, II. Structural stability, III. Priority of the octahedral and tetrahedral sites for hydrogen occupation in aluminum. IV. Anharmonic effects, V. Electronic and superconducting property, and VI. Specific heat and critical magnetic field.

\section{\label{sec:level1}I. Computational methods}
The structural optimization and thermodynamic stability of Al-based superconductor $\rm {Al_4H}$ have been systematically investigated by using the density functional theory (DFT) method as implemented in the Vienna $ab$ $initio$  simulation package (VASP)~{\color{blue}\cite{Kresse1996}}. The electron-ion potential is described by the projector augmented wave (PAW) method~{\color{blue}\cite{Blochl1994}}. To obtain optimal lattice constant, the generalized gradient approximation (GGA) of the revised Perdew-Burke-Ernzerhof (PBEsol) functional is employed to treat the exchange-correlation (XC) interactions~{\color{blue}\cite{Perdew2008}}. The kinetic energy is set to 500 eV, which is used to cut the plane wave expansion. The Monkhorst-Pack $k$-point grids of structure optimization and electronic structural calculations are chosen as 10$\times$10$\times$10 and 20$\times$20$\times$20, respectively. The convergence criteria of total energy and residual forces are smaller than $\rm {10^{-5}}$ eV and 0.001 eV/{\AA}, respectively. The $ab$ $initio$ molecule dynamics (AIMD) simulation is performed up to room temperature lasting 10 ps with a time step of 1 fs under a canonical ensemble to assess the thermodynamic stability of $\rm {Al_4H}$, in which the temperature is controlled by the Nos$\acute{\rm e}$ algorithm~{\color{blue}\cite{Nose1991}}. A $\rm {3{\times}3{\times}3}$ supercell including 135 atoms (27 H and 108 Al atoms) is constructed. Furthermore, a $\rm {3{\times}3{\times}4}$ supercell containing 180 atoms (36 H and 144 Al atoms) is also considered in order to break the symmetry of $\rm {Al_4H}$.

The electronic structures and phonon spectra of $\rm {Al_4H}$ are explored based on the DFT method by employing the QUANTUM ESPRESSO package~{\color{blue}\cite{Giannozzi2009, Giannozzi2017, Giannozzi2020}}. The GGA-PBEsol and GGA-PBE functionals are used to optimize crystal structure and calculate electronic property, respectively~{\color{blue}\cite{Perdew2008, Perdew1996}}. The kinetic energy cutoff for plane wave and charge density are set to 120 and 480 Ry. The structure is fully optimized until the force on each atom is less than $\rm {10^{-8}}$ Ry/Bohr. The charge density calculations are performed using a 18$\times$18$\times$18 $k$-point grid with a Gaussian smearing of 0.02 Ry. The dynamical matrices and vibration potentials are computed on a 6$\times$6$\times$6 $q$-point mesh within the framework of density functional perturbation theory (DFPT)~{\color{blue}\cite{Baroni2001}}. To remove the possible ambiguity due to the choice of different pseudopotentials in phonon calculations, we first choose the Optimized Norm-Conserving Vanderbilt (ONCV)~{\color{blue}\cite{Martin2015}}, PAW and ultrasoft pseudopotential (USPP) provided by \emph{PSlibrary}~{\color{blue}\cite{Andrea2014}}, a library of ultrasoft and PAW pseudopotentials, to describe the interaction between core and valence electrons in phonon spectrum calculations of aluminum, where the valence electron configurations are $2s^22p^63s^23p^1$ for ONCV and $3s^23p^1$ for PAW and USPP.

Figure~{\color{blue}\ref{Fig_S1}} (a) clearly shows that the phonon dispersions derived from ONCV, PAW and USPP are almost identical. This proves that the influence on phonon spectra from the inner $2s^22p^6$ electrons, which is much larger than that of the innermost  $1s^2$ electrons, is very small and can be ignored. Furthermore, for the purpose of comparison, we also calculate the all-electron phonon spectra of aluminum by using full-potential linearized augmented-plane wave method as implemented in the Elk code within the DFPT framework~{\color{blue}\cite{elk}}. The muffin-tin radius is set to 2.20 a.u.. Convergence is realized on a mesh of 413 $k$-points in the irreducible Brillouin zone. We find that the pseudopotential phonon calculations are in good agreement with all-electron calculations, which confirms the reliability of the pseudopotential calculations. We further carefully check the reliability of $\rm {Al_4H}$ phonon calculations by using different pseudotentials. Figure~{\color{blue}\ref{Fig_S1}} (b) demonstrates that no noticeable difference is observed among three different pseudopotentials. We thus choose the ONCV pseudopotential to treat the interactions between core and valence electrons in our following calculations. Moreover, the acoustic sum rule (ASR) is employed to process the acoustic modes at the center of the Brillouin zone ($q=0$). Two commonly-used ASRs including ``crystal" and ``simple", as well as ``no" ASR, are adopted to plot the phonon dispersions of $\rm {Al_4H}$, as shown in Fig.~{\color{blue}\ref{Fig_S1}} (c). It can be seen that the three methods show almost identical dispersions, which indicates that the ASR is unimportant. Consequently, the ASR ``crystal" is chosen to calculate phonon spectra in this Letter.

\begin{figure}[h]
\centering
\includegraphics[width=17.2cm]{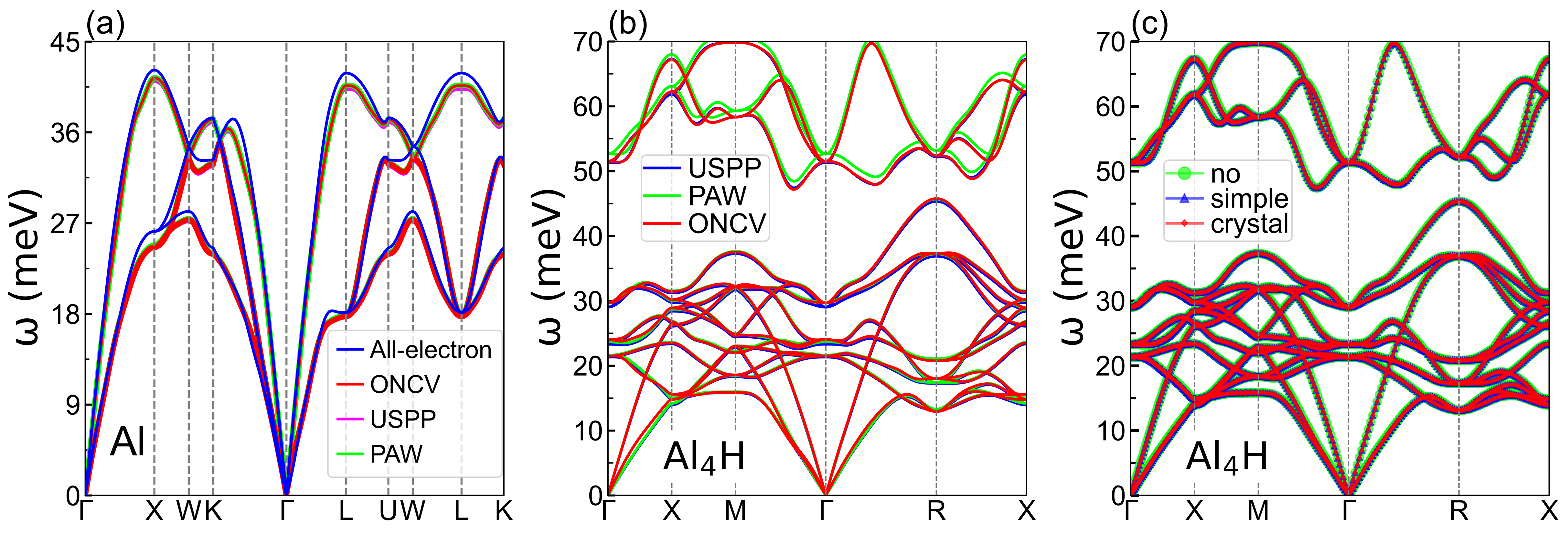}
\caption{\label{Fig_S1} Calculated phonon spectra of fcc aluminum and perovskite $\rm {Al_4H}$. (a) The phonon dispersions of aluminum are derived from ONCV, USPP and PAW pseudopotentials, respectively, where the valence electron configurations are $2s^22p^63s^23p^1$ for ONCV and $3s^23p^1$ for USPP and PAW. To check the reliability of these pseudopotentials, the all-electron phonon spectrum is also calculated carefully by using the Elk code~{\color{blue}\cite{elk}}, an all-electron code using full-potential linearized augmented plane wave (FPLAPW) method. It can be seen that the phonon spectra derived from different pseudopotentials are in good agreement with the all-electron calculations, confirming the reliability of pseudopotentials. (b) The calculated phonon spectra of $\rm {Al_4H}$ by using ONCV, USPP and PAW pseudopotentials. Little difference can be found among them, indicating that the contribution from the inner $2s^22p^6$ electrons, much larger than that of the innermost  $1s^2$ electrons, can be ignored. (c) The phonon dispersions of $\rm {Al_4H}$ are treated by using different acoustic sum rules (ASRs) including ``simple'' and ``crystal'', and without ASR ``no''. We find that the three ASR methods show almost identical results. The influence of ASR can thus be ignored.}
\end{figure}

\begin{figure}[h]
\centering
\includegraphics[width=17.2cm]{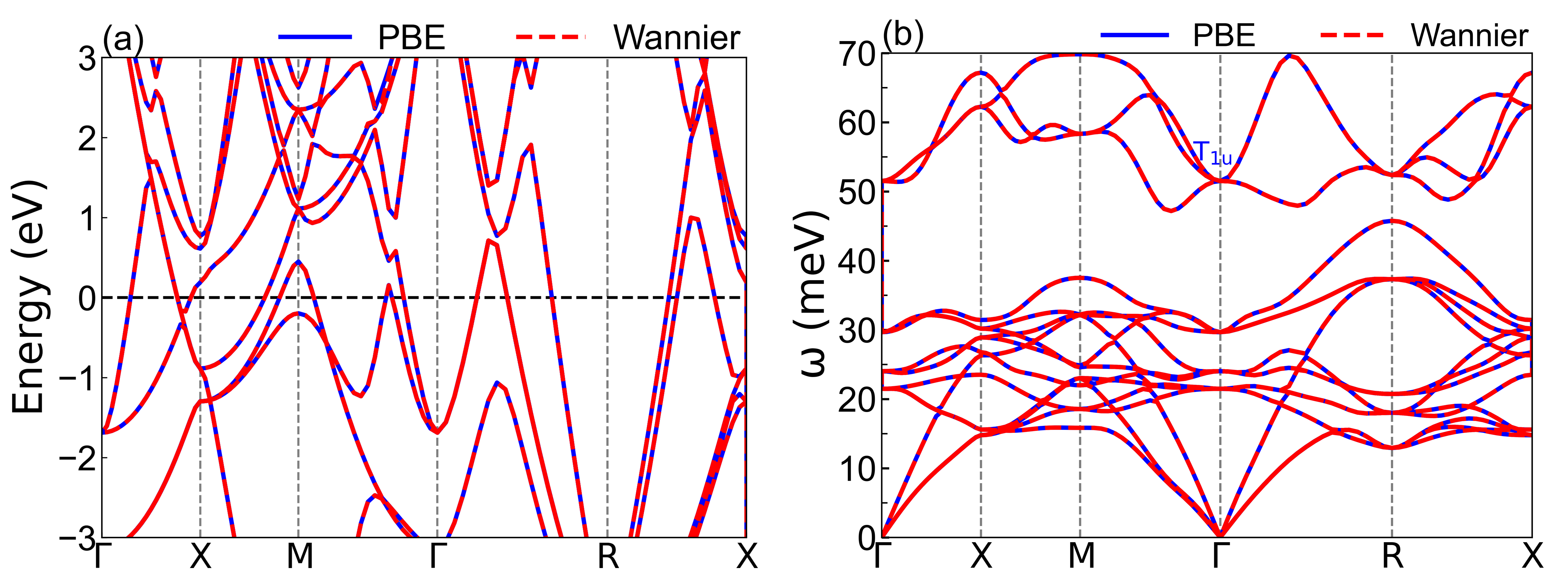}
\caption{\label{Fig_S2}~The electronic band structures (a) and phonon spectra (b) of Al-based superconductor $\rm {Al_4H}$. The blue lines are calculated by using first-principles PBE calculations, and the red dashed lines are simulated by using Wannier interpolation technology, respectively.}
\end{figure}

The electron-phonon couping (EPC) and superconducting property are calculated by using the EPW code~{\color{blue}\cite{Giustino2007, Ponce2016}} and solving the Migdal-Eliashberg (ME) equations~{\color{blue}\cite{Allen1983, Margine2013}}. The electronic wave functions are calculated using a dense $k$-point grid 18$\times$18$\times$18 within the EPW code, in order to perform the Wannier interpolation~{\color{blue}\cite{Giustino2007, Giustino2017}}. The electronic properties near the Fermi surface (FS) are described by constructing the maximally localized seventeen Wannier functions including one $s$ state of H atom, four $s$ and $p$ states for each Al atom~{\color{blue}\cite{Pizzi2020}}. The Wannier interpolated band structures and phonon spectra are shown in Fig.~{\color{blue}\ref{Fig_S2}}. The fully anisotropic ME equations can be described as follows~{\color{blue}\cite{Margine2013}},
\begin{equation}\label{equ_S1}
    \begin{aligned} Z\left(\mathbf{k}, i \omega_{n}\right)=& 1+\frac{\pi T}{N(E_{\mathrm{F}}) \omega_{n}} \sum_{\mathbf{k}^{\prime} n^{\prime}} \frac{\omega_{n^{\prime}}}{\sqrt{\omega_{n^{\prime}}^{2}+\Delta^{2}\left(\mathbf{k}^{\prime}, i \omega_{n^{\prime}}\right)}}\\& \times \delta\left(\epsilon_{\mathbf{k}^{\prime}}\right) \lambda\left(\mathbf{k}, \mathbf{k}^{\prime}, n-n^{\prime}\right), \end{aligned}
\end{equation}
\begin{equation}\label{equ_S2}
    \begin{aligned} Z\left(\mathbf{k}, i \omega_{n}\right) \Delta\left(\mathbf{k}, i \omega_{n}\right)=& \frac{\pi T}{N(E_{\mathrm{F}})} \sum_{\mathbf{k}^{\prime} n^{\prime}} \frac{\Delta\left(\mathbf{k}^{\prime}, i \omega_{n^{\prime}}\right)}{\sqrt{\omega_{n^{\prime}}^{2}+\Delta^{2}\left(\mathbf{k}^{\prime}, i \omega_{n^{\prime}}\right)}} \\& \times \delta\left(\epsilon_{\mathbf{k}^{\prime}}\right)\left[\lambda\left(\mathbf{k}, \mathbf{k}^{\prime}, n-n^{\prime}\right)-\mu^{*}\right]. \end{aligned}
\end{equation}
Here $Z({\mathbf{k}}, i{\omega}_n)$ is the mass renormalization function and ${\Delta}({\mathbf{k}}, i{\omega}_n)$ is the superconducting gap function. The $\mathbf{k}~(n)$ is the electron wave vector (band index), and ${\mu}^*$ is the semiempirical Coulomb repulsion pseudopotential, which has been set to 0.10 and 0.13 in our calculations. The anisotropic EPC strength ${\lambda}({\mathbf{k}, {\mathbf{k}^{\prime}}, n-{n}^{\prime}})$ can be calculated by using the following formula,
\begin{equation}\label{equ_S3}
    \lambda\left(\mathbf{k}, \mathbf{k}^{\prime}, n-n^{\prime}\right)=N(E_{\mathrm{F}}) \sum_{\nu} \frac{2 \omega_{\mathbf{q} \nu}}{\left(\omega_{n}-\omega_{n^{\prime}}\right)^{2}+\omega_{\mathbf{q} \nu}^{2}}\left|g_{\mathbf{k} \mathbf{k}^{\prime}}^{\nu}\right|^{2}.
\end{equation}
It is worth emphasizing that the ME equations can only be solved self-consistently along the imaginary axis at the fermion Matsubara frequencies ${\omega}_n=(2n+1){\pi}T$ ($n$ an integer) for each temperature $T$ because two equations are coupled with each other in a non-linear fashion. The numerical solutions of the ME equations for $\rm {Al_4H}$ are performed on the dense grids including $\rm {60^3}$ $\mathbf{k}$ points and $\rm {30^3}$ $\mathbf{q}$ points. The upper limit of Matsubara frequency is set to 10 times of the phonon cutoff frequency. The Dirac $\delta$ function has been represented by Lorentzian of width 50 meV for electrons. The phonon smearing parameter is carefully checked with different values of  0.05, 0.10 and 0.30 meV (see Fig.~{\color{blue}\ref{Fig_S3}}). We can find from Fig.~{\color{blue}\ref{Fig_S3}} that the superconducting gap distribution, derived from smearing parameters of 0.10 and 0.30 meV, has a negligible difference. Therefore, the Dirac $\delta$ function for phonons has been represented by Lorentzian of width 0.30 meV in the present calculations.

\begin{figure}[htp]
\centering
\includegraphics[width=10cm]{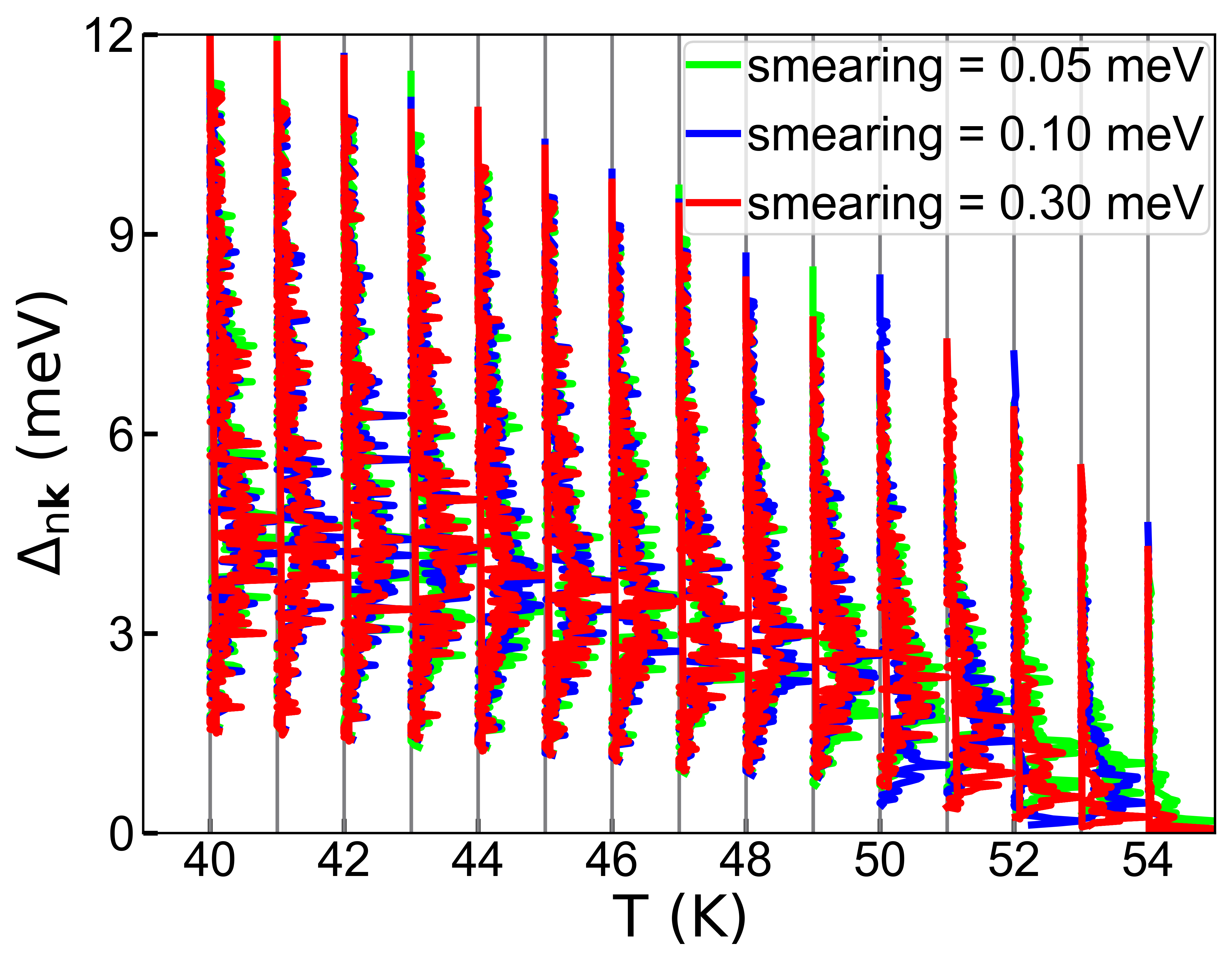}
\caption{\label{Fig_S3}Calculated superconducting gap distributions with different phonon smearing parameters of 0.05, 0.10 and 0.30 meV.}
\end{figure}

The electronic specific heat is calculated from the free energy difference between the superconducting and normal states as follows~{\color{blue}\cite{John1964, Choi2003}},
\begin{equation}\label{equ_S4}
 \begin{aligned} \Delta F=&-\pi T \sum_{n \mathbf{k} j}\left[\sqrt{\omega_{j}^{2}+\Delta_{n \mathbf{k}}^{2}\left(i \omega_{j}\right)}-\left|\omega_{j}\right|\right] \\& \times\left[Z_{n \mathbf{k}}\left(i \omega_{j}\right)-Z_{n \mathbf{k}}^{N}\left(i \omega_{j}\right) \frac{\left|\omega_{j}\right|}{\sqrt{\omega_{j}^{2}+\Delta_{n \mathbf{k}}^{2}\left(i \omega_{j}\right)}}\right] \\& \times\delta\left(\varepsilon_{n \mathbf{k}}-\varepsilon_{\mathrm{F}}\right). \end{aligned}
\end{equation}
Here, $Z^N$ represents the mass renormalization function of the normal state $N$, calculated by setting ${\Delta}$=0 in Eq.~{\color{blue}(\ref{equ_S1})}. The specific heat can be calculated from the second-order derivative of the free energy according to the following formula,
\begin{equation}\label{equ_S5}
 \Delta C(T)=-T \frac{d^{2} \Delta F}{d T^{2}}.
\end{equation}

The upper critical magnetic field $H_C$(0) can be calculated using the following equation under the condition of $T_c$/${\omega}_{\rm log}$$<$0.25~{\color{blue}\cite{Carbotte1990}},
\begin{equation}\label{equ_S6}
  \frac{\gamma T_{C}^{2}}{H_{C}^{2}(0)}=0.168\left[1-12.2\left(\frac{T_{C}}{\omega_{\log }}\right)^{2} \ln \left(\frac{\omega_{\log }}{3 T_{C}}\right)\right],
\end{equation}
where $T_c$ and ${\omega}_{\rm {log}}$ are critical temperature and logarithmic average phonon frequency, respectively. The Sommerfeld constant $\gamma$ can be calculated from the following formula,
\begin{equation}\label{equ_S7}
  \gamma=\frac{2}{3} \pi^{2} k_{\mathrm{B}}^{2}N(E_{\mathrm{F}})(1+\lambda).
\end{equation}

\section{\label{sec:level2}II. Structural stability}

\begin{figure}[h]
\centering
\includegraphics[width=14cm]{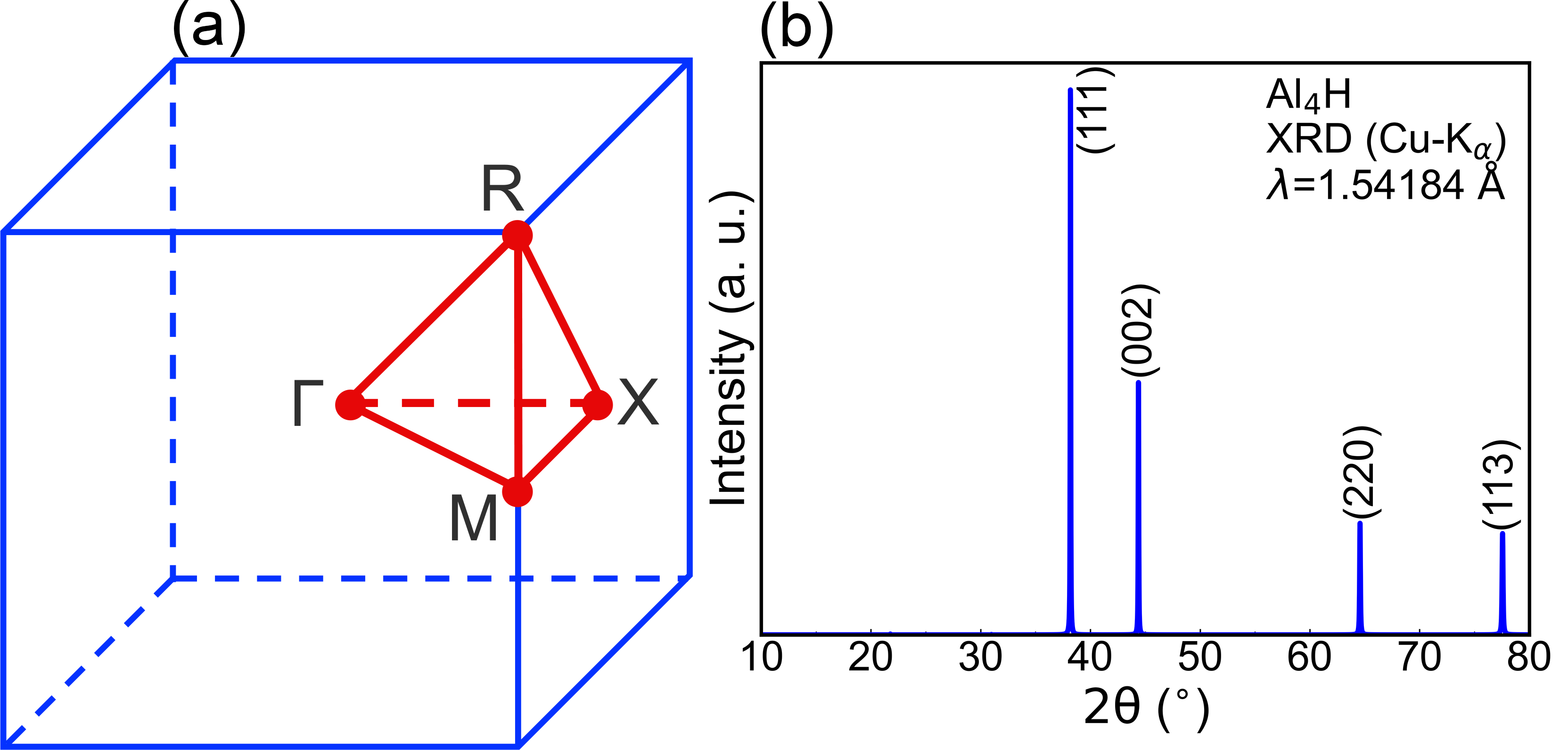}
\caption{\label{Fig_S4} (a) The first Brillouin zone of perovskite $\rm {Al_4H}$, where several high symmetry points are labeled by $\Gamma$, X, M and R, respectively. (b) The simulated XRD pattern for the optimized $\rm {Al_4H}$ from 10$^{\circ}$ to 80$^{\circ}$ by using the radiation wavelength $\lambda$=1.54184 {\AA} (Cu-K$_\alpha$). Several important peaks are marked with the corresponding Miller indices ($hkl$).}
\end{figure}

The first Brillouin zone of metal-bonded perovskite $\rm {Al_4H}$ with space group $Pm$-$3m$ (No. 221) is plotted in Fig.~{\color{blue}\ref{Fig_S4}} (a) and the simulated XRD pattern is presented in Fig.~{\color{blue}\ref{Fig_S4}} (b) by using VESTA code~{\color{blue}\cite{Momma2011}}. Table~{\color{blue}\ref{tables1}} summarizes the lattice constant, atomic distance, Wyckoff position, Bader charge and conventional cell volume of $\rm {Al_4H}$. For comparison, the corresponding results of fcc aluminum are also given. Compared with aluminum, the calculated lattice constant of $\rm {Al_4H}$ by using the advanced GGA-PBEsol XC functional has only a small increase from 4.02 to 4.08 ${\rm \AA}$, which has also been confirmed by experiments of implanting hydrogen into fcc aluminum~{\color{blue}\cite{Wei1987_2}}. Moreover, the Bader charge analysis clearly reveals that the body-centered H atom obtains about 1.41 $e$ from its neighboring Al atoms, especially the face-centered Al atoms, because the electronegativity of H atom is much larger than that of Al atom.

\begin{table*}[h]
\caption{\label{tables1}~The calculated lattice constant $a$=$b$=$c$, atomic distance $\rm Al_{(face~center)}$-$\rm H_{(body~center)}$ ($\rm Al_{(vertex)}$-$\rm H_{(body~center)}$) between the body-centered H atom and face-centered (vertex) Al atom, Wyckoff position, Bader charge and the conventional-cell volume of perovskite $\rm {Al_4H}$. The corresponding results of fcc aluminum are also given.}
\begin{ruledtabular}
\begin{tabular}{ccccllcc}
 \multicolumn{1}{c}{Material}&\multicolumn{1}{c}{$a$ ($\rm {\AA}$)}&\multicolumn{2}{c}{Atomic distance (\AA)}&\multicolumn{1}{c}{Atom}&\multicolumn{1}{l}{Wyckoff position}&\multicolumn{1}{c}{Bader charge ($e$)}&\multicolumn{1}{c}{Volume ($\rm {\AA}^3$)}\\
 \colrule
 & & $\rm Al_{(face~center)}$-$\rm H_{(body~center)}$ & 2.04 & Al (vertex) & 1a (0, 0, 0) & 0.03 & \\
 \cmidrule(r){5-7}
$\rm {Al_4H}$ (perovskite) & 4.08 & & & H (body center) & 1b (1/2, 1/2, 1/2) & -1.41 & 67.92 \\
 \cmidrule(r){5-7}
 & &  $\rm Al_{(vertex)}$-$\rm H_{(body~center)}$ & 3.53 & Al (face center) & 3c (0, 1/2, 1/2) & 3$\times$0.46 & \\
 \colrule
 Al (fcc) & 4.02 & \multicolumn{2}{c}{-} & Al & 4a (0, 0, 0) & \multicolumn{1}{c}{-} & 64.96 \\
\end{tabular}
\end{ruledtabular}
\end{table*}

It is well-known that the structural stability of perovskite $\rm {Al_4H}$, implanted hydrogen into fcc aluminum, is a foundation of its superconductivity and application. We thus comprehensively check its stability by using various methods including crystal structure analysis, theoretical calculations and experiments of hydrogen implantation in aluminum single crystals. From the viewpoint of crystal structure, the most stable structure of metal aluminum is the fcc lattice with two highly competitive interstitial sites, i.e., octahedral ($\rm O_h$) and tetrahedral ($\rm T_d$) positions. These interstitial sites can be occupied by H atom due to its smallest radius, which has been confirmed by previous experiments~{\color{blue}\cite{Bugeat1976, Bugeat1979, Schluter1993, McLellan1983, Papp1977, Lamoise1975, Hashimoto1983, Ambat1996}}. Our designed metal-bonded perovskite $\rm {Al_4H}$ can also be equivalently regarded as an ultrashort-period $\rm {(Al_4)_1H_1}$ superlattice, constructed by a close-packed Al-monolayer, identical to the one in fcc aluminum, and hexagonal H-monolayer with area density $\sigma_{\rm H}=\sigma_{\rm Al}/{\rm 4}$, similar to the one in hexagonal close-packed (hcp) hydrogen single crystal~{\color{blue}\cite{Mills1965}}. The body-centered H atom in $\rm {Al_4H}$, occupied “sphere interstice” of Al atom lattice, is naturally fixed by the stable skeleton of Al atoms without any extrinsic pressure. Similar to the metal-bonded aluminum, the perovskite $\rm {Al_4H}$, combined by the Al-H metallic bonds (see Fig.~{\color{blue}2}~(b)), also ensures that the lattice, immersed in an almost uniform electron sea, is used to fix H atom into interstitial site $\rm O_h$ due to its lower energy than $\rm T_d$ (see Table~{\color{blue}\ref{tables2}}). The structural stability of few-hydrogen metal-bonded perovskite $\rm {Al_4H}$ is quite different from the previous known multi-hydrogen superconducting hydrides under ultrahigh pressures and perovskite semiconductors combined with ionic and covalent bonds because both the skeleton of the stable metal Al lattice and electron sea can be used to fix H atoms in few-hydrogen perovskite $\rm {Al_4H}$.

In order to certify the stability of perovskite $\rm {Al_4H}$ from theory, we firstly calculate its formation energy defined by $E_{\rm f} $=$ (E_{\rm {Al_4H}} $-$ E_{\rm {H}} $-$ 4E_{\rm {Al}})/5$, in which $E_{\rm {Al_4H}}$ is the total energy of $\rm {Al_4H}$. The $E_{\rm {Al}}$ represents the energy of per Al atom in its corresponding single crystal phase. The $E_{\rm {H}}$ is the energy of per H atom in the hcp hydrogen single crystal phase~{\color{blue}\cite{Mills1965}}. Our calculated $E_{\rm f}$ is -0.29 eV/atom, indicating the structural stability of $\rm {Al_4H}$. To further confirm perovskite Al$_4$H having the minimum energy, we carefully investigate the energy dependence of  Al:H=4:1 alloying configurations within the framework of perovskite lattice by using a 2$\times$2$\times$1 supercell. A total of 261 supercells were generated by using $disorder$ code with enumerated method~{\color{blue}\cite{Lian2020}}. Figure~{\color{blue}\ref{Fig_S5}} (a) shows that there are fourteen different structures with nearly identical total energy without zero-point energy (ZPE) correction. To distinguish these configurations carefully, we further calculate their total energies including ZPE correction due to the lightest H atom with large amplitude. We find from Fig.~{\color{blue}\ref{Fig_S5}} (b) that the last configuration, named as ``261", i.e., the perovskite Al$_4$H structure that we designed, has the lowest energy among all configurations. The total energy of the other configuration, named as ``190", with H atom also occupying O$\rm _h$ site, has almost the same total energy as perovskite Al$_4$H. It is clear that H atom at the octahedral interstice O$\rm _h$ in fcc aluminum has the lowest energy, indicating the structural stability of perovskite Al$_4$H.

Let us now further perform the AIMD simulations to assess the thermodynamic stability of H atom located at $\rm {T_d}$ and $\rm {O_h}$ sites at different temperatures. Our calculation results are presented in Figs.~{\color{blue}\ref{Fig_S6}} and~{\color{blue}\ref{Fig_S7}}. To guarantee the reliability of the AIMD simulation, the 3${\times}$3${\times}$3 supercell of $\rm {Al_4H}$ with H atom located at $\rm {T_d}$ site is constructed, in which 135 atoms with 27 H and 108 Al atoms are included. The results demonstrate that the total energy exhibits an obvious decrease from 0 to 2 ps, indicating that H atom is very hard to be fixed at its initial $\rm {T_d}$ position. The final structure snapshots clearly show that some H atoms migrate from $\rm {T_d}$ to $\rm {O_h}$ sites from 50 to 300 K (see Fig.~{\color{blue}\ref{Fig_S6}}), which clearly demonstrates that H atoms prefer to occupy $\rm {O_h}$ sites in fcc Al lattice at $T\neq$0 K. The thermodynamic stability of perovskite $\rm {Al_4H}$ with H atom at $\rm {O_h}$ site is also investigated by the AIMD simulation at 300 K. To break the structural symmetry, both the 3${\times}$3${\times}$3 and 3${\times}$3${\times}$4 (36 H and 144 Al atoms) supercells are constructed (see Fig.~{\color{blue}\ref{Fig_S7}}). The results show that the total energy fluctuation is very small and all atoms slightly vibrate around their equilibrium positions. More specifically, H atoms are always at their equilibrium positions, avoiding to form $\rm {H_2}$ molecules due to migration of H atoms and guaranteeing structural stability of perovskite $\rm {Al_4H}$.

We further calculate the total energy curve as a function of H atom displacement along the [111] diagonal direction $\rm {T_d}$-$\rm {O_h}$-$\rm {T_d}$, i.e., the minimum energy path, where both the lattice constant and atomic position are fully relaxed and the zero-point energy is included. It is seen that the total energy for H atom at $\rm {O_h}$ site is lower than that at the $\rm {T_d}$ site by 0.09 eV, indicating that H atom prefers to occupy the $\rm {O_h}$ position in fcc aluminum (see Fig.~{\color{blue}\ref{Fig_S8}}). The migration of H atom from $\rm {O_h}$ to $\rm {T_d}$ site can be suppressed because the high energy barrier of 0.18 eV exists. The opposite migration process for H atom from $\rm {T_d}$ to $\rm {O_h}$ site becomes possible due to the low energy barrier of 0.09 eV. Moreover, we also carefully calculate the phonon spectra of perovskite $\rm {Al_4H}$ (see Fig.~{\color{blue}3}~(b), Figs.~{\color{blue}\ref{Fig_S1}} (b) and (c) and~{\color{blue}\ref{Fig_S2}}~(b)) to testify its dynamic stability, in which three different pseudopotentials are adopted to remove the ambiguity of pseudopotential choice. The different ASRs including ``crystal" and ``simple", as well as ``no" ASR, are also checked. Figure~{\color{blue}3}~(b) and Figs.~{\color{blue}\ref{Fig_S1}} (b) and (c) and~{\color{blue}\ref{Fig_S2}}~(b) clearly show that no imaginary frequency phonon modes exist, guaranteeing favorable dynamic stability of perovskite $\rm {Al_4H}$. Moreover, we also adopt much denser $k$-point mesh 24$\times$24$\times$24 and $q$-point mesh 12$\times$12$\times$12 as well as different XC functionals (PBE, PBEsol and PW91) to calculate the phonons exactly at $\Gamma$-point. All of our calculations clearly demonstrate that there is still no imaginary frequency at $\Gamma$-point. At the same time, our calculations also clearly demonstrate that the phonon dispersions at $\Gamma$-point are convergent. Therefore, the dynamical stability of Al-based superconductor Al$_4$H has been fully affirmed.

\begin{figure}[htp]
\centering
\includegraphics[width=13cm]{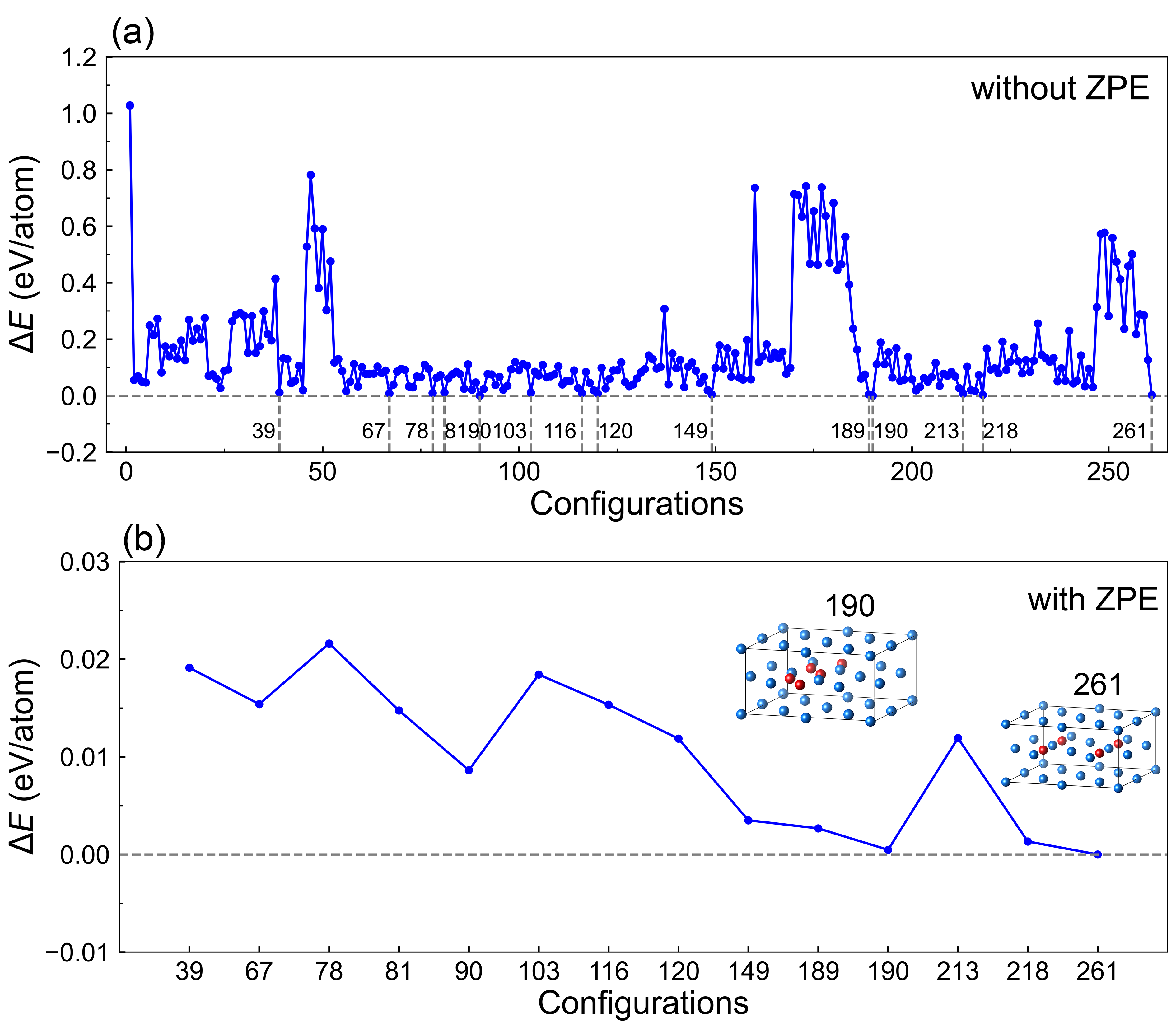}
\caption{\label{Fig_S5}Total energy dependence of Al:H=4:1 alloying configurations within the framework of perovskite lattice with a total of 261 2$\times$2$\times$1 supercells generated by using $disorder$ code with enumerated method~{\color{blue}\cite{Lian2020}}. (a) Calculated total energies without zero-point energy (ZPE) correction. The minimum energy among all of 261 different configurations is set to zero. It can be seen that there are fourteen configurations, numbered as 39, 67, 78, 81, 90, 103, 116, 120, 149, 189, 190, 213, 218 and 261,  have almost identical total energy. (b) Similar to (a), but for the aforementioned fourteen different structures including ZPE correction. The insets are the lattice structures of ``190" and ``261'' configurations, where the blue and red balls represent Al and H atoms, respectively. The perovskite Al$_4$H, i.e., the ``261'' configuration, with H atom at the octahedral interstice O$\rm _h$ in fcc aluminum, has the lowest energy.}
\end{figure}

\begin{figure}[htp]
\centering
\includegraphics[width=16cm]{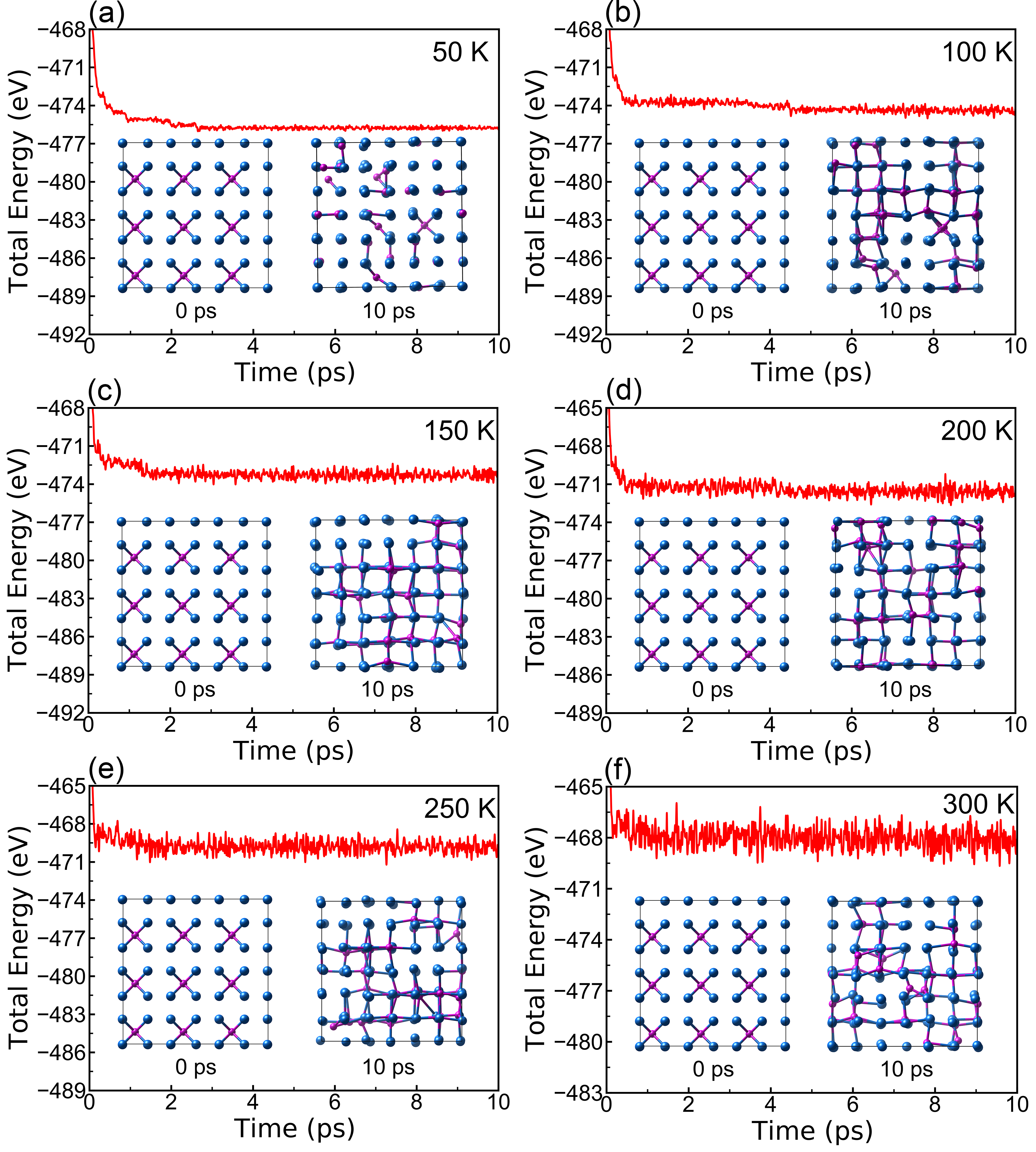}
\caption{\label{Fig_S6} Calculated total energy of $\rm {Al_4H}$ with H atom located at  tetrahedral interstice $\rm {T_d}$ as a function of simulation time during AIMD simulation from 50 to 300 K (a-f), in which the 3${\times}$3${\times}$3 supercell with 135 atoms (27 H and 108 Al atoms) is constructed. The structure snapshots at 0 and 10 ps are also given. The results demonstrate that the total energy exhibits an obvious decrease from 0 to 2 ps. This indicates that it is very difficult to hold H atom at its initial $\rm {T_d}$ position. The final structure snapshots show a similar tendency from  50 to 300 K. The migration of some H atoms from $\rm {T_d}$ to $\rm {O_h}$ sites can be observed clearly during the AIMD simulation at different temperatures, which demonstrates that H atoms prefer to occupy $\rm {O_h}$ sites with the lowest energy in fcc Al lattice (see Table~{\color{blue}\ref{tables2}}), rather than $\rm {T_d}$ sites. Moreover, the average energy shown in (f) is -468 eV for the 3${\times}$3${\times}$3 supercell at 300 K, which is larger than -469.5 eV given in Fig.~{\color{blue}\ref{Fig_S7}} (a) for H atom at the octahedral interstice $\rm {O_h}$ site.}
\end{figure}

\begin{figure}[htp]
\centering
\includegraphics[width=17.2cm]{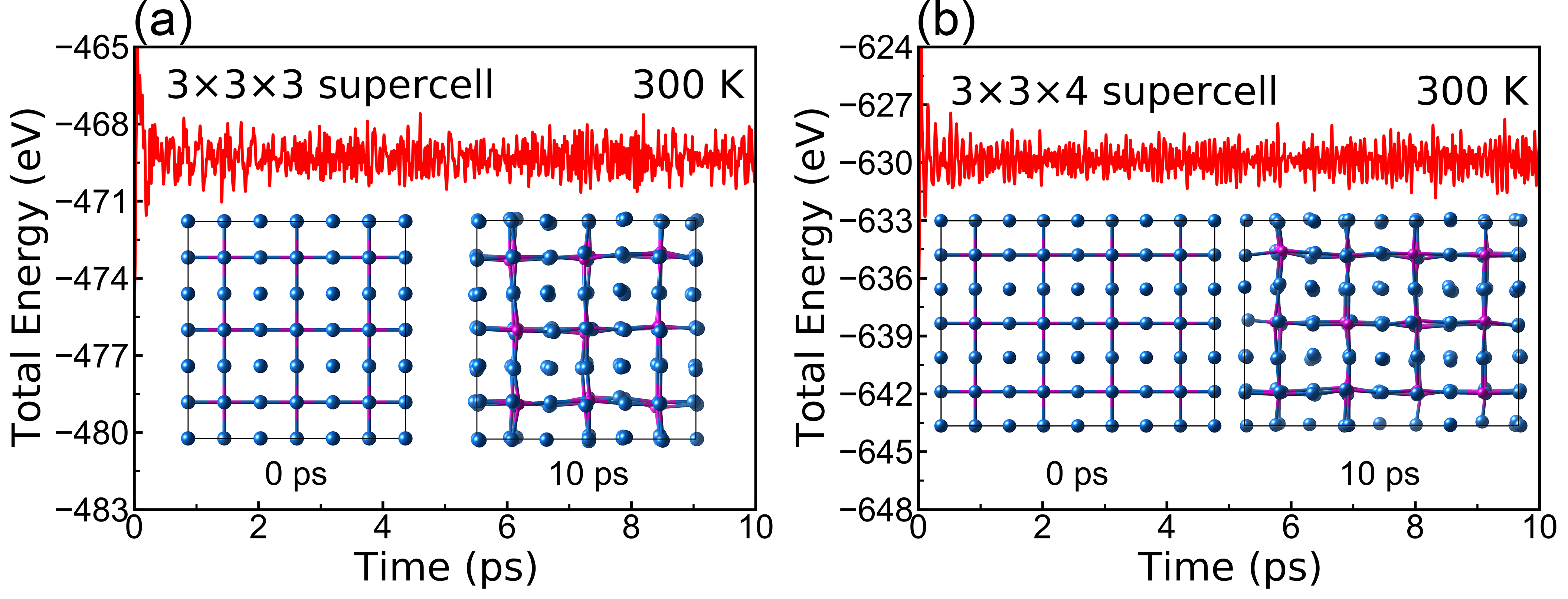}
\caption{\label{Fig_S7}~Time dependence of the total energy of perovskite $\rm {Al_4H}$ with H atom at the octahedral interstice $\rm {O_h}$ during AIMD simulation at 300 K. (a) The 3${\times}$3${\times}$3 supercell, containing 135 atoms (27 H and 108 Al atoms) with an average energy of -469.5 eV, and (b) 3${\times}$3${\times}$4 supercell, including 180 atoms (36 H and 144 Al atoms), are used to break the structural symmetry. The insets are structure snapshots at 0 and 10 ps. The results show that the total energy fluctuation is very small and all atoms slightly vibrate around their equilibrium sites. More specifically, the H atom is always at its equilibrium position, avoiding to form $\rm {H_2}$ molecule due to migration of H atoms and guaranteeing structural stability of perovskite $\rm {Al_4H}$.}
\end{figure}

\begin{figure}[htp]
\centering
\includegraphics[width=10.5cm]{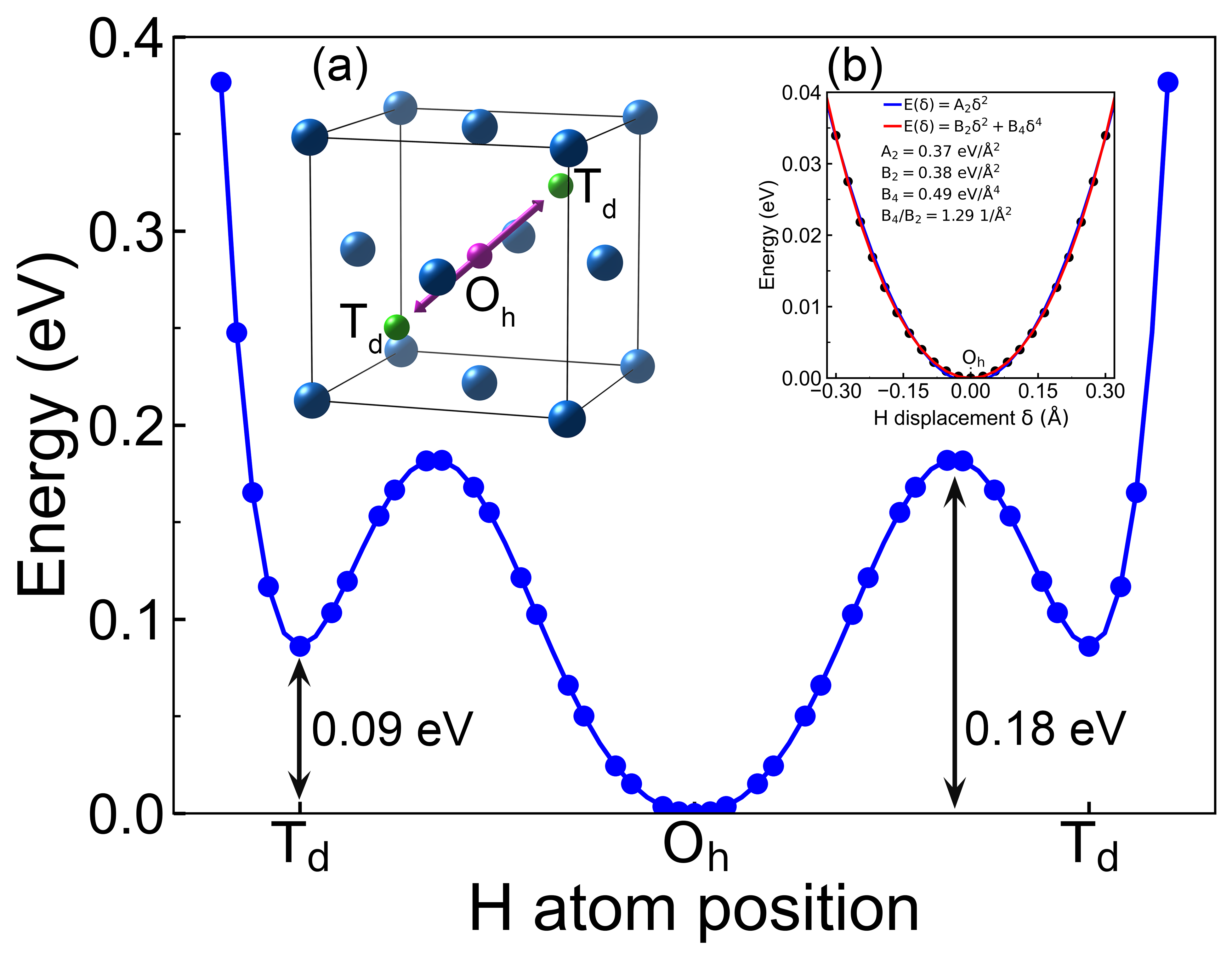}
\caption{\label{Fig_S8}~The calculated total energy of Al$_4$H conventional cell as a function of the H atom's position along the [111] diagonal direction $\rm T_d$-$\rm O_h$-$\rm T_d$ (see inset (a)), i.e., the minimum energy path, in which the zero-point energy is included and the lattice is relaxed. The inset (b) shows the variation of the total energy for the H atom microvibration near $\rm O_h$ with a small displacement $\delta$. It clearly indicates that the potential well can be fit neatly to E($\delta$)=B$_2$$\delta^2$+B$_4$$\delta^4$ with a small ratio of  B$_4$/B$_2$=1.29 1/{${\rm \AA}^2$}. Moreover, the difference between the fourth power fitting and parabolic fitting is also negligible. Therefore, the local potential around $\rm O_h$ is weakly anharmonic in Al-based superconductor Al$_4$H.}
\end{figure}

From experimental point of view, hydrogen implantation and diffusion in fcc aluminum, with two natural interstitial sites $\rm O_h$ and $\rm T_d$ occupied by the smallest radius H atom, have been investigated previously. By means of the low temperature hydrogen implantation method, Bugeat $et$ $al.$ realized an effective hydrogen implantation in aluminum at 10 and 150 keV under 25-400 K and studied lattice localization and trapping of H atoms~{\color{blue}\cite{Bugeat1976, Bugeat1979}}. Hydrogen diffusion measurements were performed by applying the potentiostatic and galvanostatic permeation and the current pulse method using the electrochemical double cell technique~{\color{blue}\cite{Schluter1993}}. Hydrogen diffusion is also investigated by analyzing the kinetic behavior predicted for solutions of hydrogen in aluminum~{\color{blue}\cite{McLellan1983}} and by the desorption method~{\color{blue}\cite{Papp1977}}. Wei $et$ $al.$ proved that H atoms can be introduced into the thin crystalline aluminum films at 10-20 K with the overall long-range order and the fcc Al lattice is always stable even if the atomic ratio H:Al is about 1:1~{\color{blue}\cite{Wei1987_2}}. Moreover, Lamoise $et$ $al.$ reported that H atoms can be implanted into aluminum thin films with implantation energy of 10 keV at temperature below 6 K at a high average concentration of H:Al ratio up to 2:1 by using the Orsay ion implantor~{\color{blue}\cite{Lamoise1975}}. Hashimoto $et$ $al.$ also performed an effective hydrogen implantation in aluminum crystals by gas phase charging in the temperature range 300-400  $^{\circ}$C and by electrochemical charging at room temperature~{\color{blue}\cite{Hashimoto1983}}. As stated above, H atoms can be implanted into fcc aluminum with a high atomic ratio H:Al about 2:1. The stability of fcc Al lattice implanted by H atoms, occupied the interstitial site $\rm O_h$ or $\rm T_d$, has been confirmed by many previous experiments~{\color{blue}\cite{Bugeat1976, Bugeat1979, Schluter1993, McLellan1983, Papp1977, Lamoise1975, Hashimoto1983, Ambat1996}}. This clearly indicates that perovskite $\rm {Al_4H}$ with H atom at the body center $\rm O_h$ due to its lower energy than $\rm T_d$ site  (see Fig.~{\color{blue}\ref{Fig_S8}} and Table~{\color{blue}\ref{tables2}}) is the most stable phase under the conditions of fcc Al lattice and easily realized low H concentration of H:Al=1:4, much smaller than H:Al=2:1 as reported in experiments~{\color{blue}\cite{Lamoise1975}}.

Based on the structural characteristics of $\rm {Al_4H}$, the aforementioned theoretical analyses about its stability and the experimental hydrogen implantation in fcc aluminum crystals, we can suggest five potential experimental synthesis schemes to prepare perovskite $\rm {Al_4H}$ by two steps, i.e., (1) hydrogen implantation and (2) hydrogen diffusion in fcc aluminum crystals. Firstly, H atoms can be implanted into aluminum by using the low temperature hydrogen implantation method as adopted previously~{\color{blue}\cite{Lamoise1975}}. Secondly, H atoms can also be introduced into aluminum by gas phase charging of H atoms under ambient pressure at 300 $^{\circ}$C~{\color{blue}\cite{Hashimoto1983}}. Thirdly, by means of the electrochemical charging at room temperature~{\color{blue}\cite{Hashimoto1983}}, H atoms can also be implanted into fcc aluminum. Fourthly, $\rm {Al_4H}$ may also be synthesized via a direct reaction of aluminum and hydrogen at high pressure, i.e., pressure-driven formation, in which H atoms are pressed into aluminum, similar to the synthesis of lanthanum hydrides~{\color{blue}\cite{Drozdov2019}}. After H atoms are implanted into aluminum by using the above methods, both the $\rm {O_h}$ and $\rm {T_d}$ sites will be occupied by H atoms. Considering that the low energy barrier (0.09 eV) for H atom migration from $\rm {T_d}$ to $\rm {O_h}$ site and the lowest energy of the $\rm {O_h}$ site (Fig.~{\color{blue}\ref{Fig_S8}} and Table~{\color{blue}\ref{tables2}}), and AIMD simulations (Figs.~{\color{blue}\ref{Fig_S6}} and~{\color{blue}\ref{Fig_S7}}), we can confirm that H atoms prefer to occupy $\rm {O_h}$ sites. Therefore, we can expect that H atoms at $\rm {T_d}$ sites may overcome the low energy barrier and migrate to $\rm {O_h}$ sites in fcc Al lattice to form the favorable perovskite $\rm {Al_4H}$ under ambient pressure and at room temperature after a period of time, for example, two weeks~{\color{blue}\cite{Drozdov2019}}. Finally, perovskite $\rm {Al_4H}$ viewed along the [111] direction, equivalent to an ultrashort-period $\rm {(Al_4)_1H_1}$ superlattice, can also be expected to achieve its controllable preparation directly on Al substrate by adopting plasma-assisted molecular beam epitaxy (MBE) with k-cell source to steam Al and hydrogen plasma source to steam H, accurately controlling H:Al ratio in the excess hydrogen environment, and adjusting the substrate temperature at low temperature, for example, the liquid-nitrogen or liquid-helium temperature, and using the sub-monolayer digital-alloying technique~{\color{blue}\cite{Jmerik2009}}.

\section{\label{sec:level3}III. Priority of the octahedral and tetrahedral sites for hydrogen occupation in aluminum}

\begin{table*}[h]
\caption{\label{tables2}~The calculated lattice constant ($a$), total energy ($\rm {E_{total}}$), zero-point energy (ZPE) and total energy difference $\rm {{\Delta}E}$=$\rm {E_{total}(O_h)}$-$\rm {E_{total}(T_d)}$ with and without ZPE contribution by using different XC functionals.}
\begin{ruledtabular}
\begin{tabular}{ccccccccc}
 \multicolumn{1}{c}{XC type}&\multicolumn{2}{c}{$a$ ($\rm {\AA}$)}&\multicolumn{2}{c}{$\rm {E_{total}}$ (eV)}&
 \multicolumn{2}{c}{ZPE (eV)}&\multicolumn{2}{c}{$\rm {\Delta}E~(meV/atom)$}\\
   & $\rm {O_h}$ & $\rm {T_d}$ & $\rm {O_h}$ & $\rm {T_d}$ & $\rm {O_h}$ & $\rm {T_d}$ & without ZPE & with ZPE\\
   \hline
    \multicolumn{9}{c}{LDA} \\ \hline
    PZ        & 4.05 & 4.07 & -19.61 & -19.66 & 0.20 & 0.29 & 10.00 & -8.00 \\
    VWN5 & 4.05 & 4.07 & -19.56 & -19.61 & 0.20 & 0.29 & 10.00 & -8.00 \\
    HL       & 4.05 & 4.07 & -21.44 & -21.49 & 0.20 & 0.29 & 10.00 & -8.00 \\
    WI       & 4.11 & 4.13 & -16.87 & -16.93 & 0.18 & 0.27 & 12.00 & -6.00 \\
    \hline
     \multicolumn{9}{c}{GGA} \\
     \hline
     PW91 & 4.11 & 4.13 & -17.35 & -17.46 & 0.15 & 0.26 & 22.00 & 0.00 \\
     PBE   & 4.10 & 4.13 & -17.56 & -17.64 & 0.18 & 0.27 & 16.00 & -2.00 \\
     revPBE & 4.11 & 4.14 & -16.77 & -16.86 & 0.17 & 0.27 & 18.00 & -2.00 \\
     RPBE & 4.13 & 4.15 & -16.58 & -16.67 & 0.17 & 0.26 & 18.00 & 0.00 \\
     PBEsol & 4.08 & 4.10 & -18.98 & -19.02 & 0.20 & 0.28 & 8.00 & -8.00 \\
     AM05 & 4.07 & 4.09 & -18.64 & -18.64 & 0.22 & 0.29 & 0.00 & -14.00 \\
\end{tabular}
\end{ruledtabular}
\end{table*}

It is worthwhile to note that the preference of H atom at two highly competitive interstitial sites ($\rm O_h$ or $\rm T_d$) in fcc aluminum has long been controversial. Based on DFT calculations, Popovic $et$ $al.$~{\color{blue}\cite{Popovic1974}} found that the energy of the H atom at the $\rm O_h$ site in aluminum is lower than that at the $\rm T_d$ site by 0.14 eV and suggested that the H atom in aluminum favors the $\rm O_h$ site, which has been confirmed experimentally by Mclellan~{\color{blue}\cite{McLellan1983}}. On the contrary, Bugeat $et$ $al.$ investigated hydrogen implantation in fcc metals and inferred that the H atom occupies the $\rm T_d$ site in fcc aluminum~{\color{blue}\cite{Bugeat1979}}. Wolverton $et$ $al.$ theoretically investigated the priority of the H atom occupied the $\rm O_h$ or $\rm T_d$ interstitial sites in aluminum by using GGA functional, in which both the atomic site optimization and zero-point vibration were considered~{\color{blue}\cite{Wolverton2004}}. Unfortunately, they only optimized all atomic positions without full relaxation of the crystal lattice. They found that the H atom in aluminum is slightly energetically preferred at the $\rm T_d$ position.

To accurately clarify the interstitial priority of H atom in fcc aluminum, we fully relax lattice constants and all atomic positions for H atom located at $\rm O_h$ and $\rm {T_d}$ sites in Al$_4$H, in which the advanced PBEsol~{\color{blue}\cite{Perdew2008}} and AM05~{\color{blue}\cite{Armiento2005}} together with the other accurate functionals are used and the zero-point energy (ZPE) contribution is considered. The calculated lattice constants for the H atom at $\rm O_h$ and $\rm {T_d}$ sites by using various XC functionals have different values (see Table~{\color{blue}\ref{tables2}}), which means that lattice constants should be fully optimized. Table~{\color{blue}\ref{tables2}} also lists the calculated ZPE and total energy difference $\rm {{\Delta}E}$=$\rm {E}_{\rm total}(\rm {O}_{\rm h})$-$\rm {E}_{\rm total}(\rm {T}_{\rm d})$ with and without the ZPE contribution for both sites. If ignoring ZPE, we find that H atom energetically prefers the $\rm {T_d}$ site with lower energy than $\rm O_h$, which is in agreement with previous calculations~{\color{blue}\cite{Wolverton2004}}. However, the zero-point vibrations are important and should be included because of the high vibrational frequency of the H atom with the lightest mass. Our calculations indicate that the $\rm {O_h}$ site has lower ZPE than the $\rm {T_d}$ site, revealing that the H atom at the $\rm {T_d}$ site exhibits a higher vibrational frequency than that at the $\rm {O_h}$ site, which is consistent with the previous calculations~{\color{blue}\cite{Wolverton2004}}. As stated above, we find that the tetrahedral site $\rm T_d$ has the lowest energy if ignoring zero-point vibration. On the contrary, the octahedral site $\rm O_h$ has the lowest energy and becomes the preference site for the H atom occupation in fcc aluminum if including the important ZPE correction, which is consistent with previous theoretical calculations~{\color{blue}\cite{Popovic1974}}. Please refer to Table~{\color{blue}\ref{tables2}} for details. Moreover, we can also find from AIMD simulations of Figs.~{\color{blue}\ref{Fig_S6}} (f) and~{\color{blue}\ref{Fig_S7}} (a) that the average energy for the H atom at the tetrahedral site $\rm T_d$  is -468 eV for the 3${\times}$3${\times}$3 supercell at 300 K, which is larger than -469.5 eV for the H atom at the octahedral interstice site $\rm {O_h}$. Based on the exact DFT calculations and AIMD simulations at 300 K, we thus solve this long-running debate and prove that the few-hydrogen binary metal-bonded perovskite Al$_4$H with H atom at $\rm O_h$ site is the most stable structure under the preconditions of fcc aluminum and Al:H=4:1.

\section{\label{sec:level4}IV. Anharmonic effects}

\begin{figure}[htp]
\centering
\includegraphics[width=17cm]{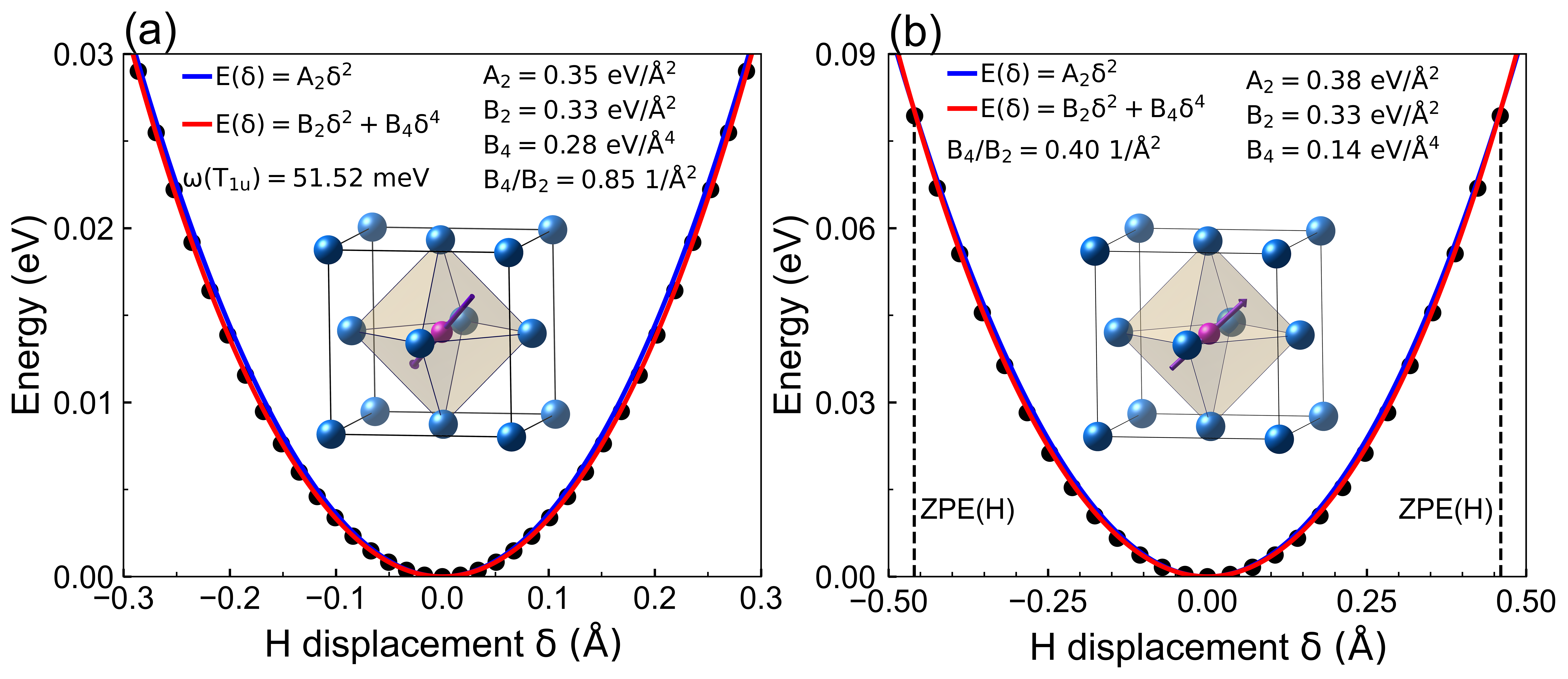}
\caption{\label{Fig_S9}Energy curves as a function of H atom displacement for (a) the zone-center frozen-phonon mode T$_{\rm 1u}$ and (b) the frozen-phonon mode along the [111] diagonal direction, i.e., the minimum energy path. It can be easily find that the potential well can be fit neatly to E($\delta$)=B$_2$$\delta^2$+B$_4$$\delta^4$ with a small ratio of (a) B$_4$/B$_2$=0.85 and (b) 0.40 1/{${\rm \AA}^2$}, respectively. The local potentials are thus weakly anharmonic in Al$_4$H due to small B$_4$/B$_2$ and minor difference between the fourth power fitting and parabolic fitting.}
\end{figure}

The anharmonic effect, caused by the light atomic mass and large displacement, represents an important quantum contribution from the kinetic term of the nuclear Hamiltonian to the energy and phonon frequency, which directly affects many physical properties of materials, such as thermal expansion, thermodynamical stability, transport characteristics, superconducting property and so on. Generally, atoms in solids vibrate around their equilibrium sites. If atomic vibration exhibits small amplitude, it can be well described by the harmonic approximation. Otherwise, anharmonicity will play an important role in solids constituted by the elements with light atomic mass, such as multi-hydrogen high-temperature superconductors because of the lightest atomic mass of the H atom with a large amplitude and high concentration. Therefore, anharmonic correction is always employed to explain the discrepancy of critical temperature between the classical harmonic approximation calculations and experimental observations in multi-hydrogen superconductors such as palladium hydrides~{\color{blue}\cite{Errea2013}}, sulfur hydrides~{\color{blue}\cite{Errea2015}} and lanthanum hydrides~{\color{blue}\cite{Errea2020}}. Here, we would like to explore the anharmonic effects in perovskite $\rm {Al_4H}$ from two aspects. On the one hand, the small hydrogen concentration in our designed $\rm {Al_4H}$ structure at ambient pressure, quite different from the widely investigated multi-hydrogen superconductors under high pressure, provides a very small quantum contribution to nuclear vibrations. Furthermore, the Bader charge analysis shows that H atom obtains about 1.41 $e$ from its neighboring Al atoms in $\rm {Al_4H}$ (see Table~{\color{blue}\ref{tables1}}), which further enlarges its effective radius, squeezes its space and suppresses its vibration. We also note that the anharmonic effects mainly correct some abnormal phonon dispersions in multi-hydrogen compounds, for example, the serious softening of phonon spectra in $\rm {H_3S}$ at 200 GPa, $\rm {LaH_{10}}$ under 264 GPa and $\rm {AlH_3}$ at 89-137 GPa derived from the harmonic approximation~{\color{blue}\cite{Errea2015, Errea2020, Hou2021}}. However, our calculated phonon dispersions of $\rm {Al_4H}$ (see Fig.~{\color{blue}3}~(b) and Figs.~{\color{blue}\ref{Fig_S1}} (b) and (c) and~{\color{blue}\ref{Fig_S2}} (b)) do not exhibit remarkable phonon softening in whole Brillouin zone, indicating weak anharmonicity in few-hydrogen perovskite $\rm {Al_4H}$.

On the other hand, the atomic potential energy in solids can be easily expanded according to Taylor series as follows,
\begin{equation}\label{equ_S8}
 V(r)=V(a+{\delta})=\underbrace{V(a)+{\frac{1}{2!}}\left(\frac{d^2V}{dr^2}\right)_a{{\delta}^{2}}}_{\rm harmonic~approximation}+\underbrace{{\frac{1}{3!}}\left(\frac{d^3V}{dr^3}\right)_a{{\delta}^3}+{\frac{1}{4!}}\left(\frac{d^4V}{dr^4}\right)_a{{\delta}^4}+...}_{\rm anharmonic~approximation},
\end{equation}
where $a$ is the equilibrium lattice constant and $\delta$ is the displacement of the atomic microvibration near its equilibrium position. Obviously, the leading term is the harmonic approximation, in which the Taylor expansion in Eq.~({\color{blue}\ref{equ_S8}}) is truncated at the second order term in $\delta^2$. The high-order terms, named as the anharmonic terms, describe the anharmonic effects of the lattice vibrations. Specifically, the term in $\delta^3$ represents the asymmetry of the mutual repulsion of the atoms and the term in $\delta^4$ represents the softening of the vibration at large amplitude. According to the phonon spectra of Al-based superconductor $\rm {Al_4H}$ (Fig.~{\color{blue}3} (b) and Figs.~{\color{blue}\ref{Fig_S1}} (b) and (c) and~{\color{blue}\ref{Fig_S2}} (b)), the high frequency modes are mainly dominated by the vibrations of H atoms, we thus pay much attention to the anharmonic effect of the $\rm T_{1u}$ vibration mode at $\Gamma$ point. To evaluate its anharmonicity, we calculate the total energy as a function of H atom displacement ${\delta}$ by using the frozen-phonon method~{\color{blue}\cite{Yin1980}}, where the energy at ${\delta}$=0 ${\rm \AA}$ is chosen as the reference energy. The calculated energy value $\rm E({\delta})$ is fitted with quadratic and quartic polynomials by using the least square method, respectively,
\begin{equation}\label{equ_S9}
\begin{aligned}
 & E({\delta})={A_2}{\delta}^2,\\
 & E({\delta})={B_2}{\delta}^2+{B_4}{\delta}^4,
\end{aligned}
\end{equation}
where the term in $\delta^3$ is zero because of the octahedral symmetry around the body-centered H atom~{\color{blue}\cite{Rush1984}}. We can find from Fig.~{\color{blue}\ref{Fig_S9}} (a) that the total energy can be well fitted by a parabola up to $\rm {{\delta}/a=0.07}$, indicating the harmonic phonons even at these large displacements. The total energy curve fitted by the quartic polynomial is nearly coincident with the parabolic fitting. The ratio $\rm {B_4}$/${\rm {B_2}}$ is only 0.85 1/${\rm \AA}^2$, which further demonstrates that perovskite $\rm {Al_4H}$ is weak anharmonicity. Moreover, we also calculate the total energy as a function of H atom displacement $\rm {\delta}$ along the [111] diagonal direction by using the frozen-phonon method at a large energy truncation of ZPE (see Fig.~{\color{blue}\ref{Fig_S9}} (b)) and by fully optimizing the lattice constants (see inset (b) of  Fig.~{\color{blue}\ref{Fig_S8}}). Both the inset (b) of Fig.~{\color{blue}\ref{Fig_S8}} and Fig.~{\color{blue}\ref{Fig_S9}} (b) clearly show that parabolic fittings are in good agreement with the fourth power fittings, indicating the weak anharmonicity of the local potential around the interstitial site $\rm O_h$ in $\rm {Al_4H}$, which has also been further confirmed by the small $\rm {B_4}$/${\rm {B_2}}$ ratio of 0.40 (the frozen-phonon method) and 1.29 (the full lattice relaxation) 1/${\rm \AA}^2$. Considering that the anharmonic phonon calculation is a very time-consuming task, we thus use harmonic approximation to investigate superconducting property of $\rm {Al_4H}$ because of its weak anharmonicity in this Letter.

\section{\label{sec:level5}V. Electronic and superconducting property}

Here, we summarize the orbital-resolved Fermi surface (Fig.~{\color{blue}\ref{Fig_S10}}), the normalized distribution of the EPC strength (Fig.~{\color{blue}\ref{Fig_S11}}) and the temperature dependence of the anisotropic superconducting gap with the semiempirical Coulomb repulsion pseudopotential $\rm {{\mu}^*=0.13}$ (Fig.~{\color{blue}\ref{Fig_S12}}) in Al-based perovskite $\rm {Al_4H}$. In order to prove that our novel idea presented in Fig.~{\color{blue}1} is fruitful and feasible, we further find that, besides perovskite $\rm {Al_4H}$ investigated in this Letter, several other binary and ternary few-hydrogen metal-bonded perovskite M$_4$H (M=Ca, Cu, Rh) and AHM$_3$ (A=V, Nb, M=Al, Rh, Ca) hydrides are also phonon-mediated superconductors under ambient pressure. The superconducting transition temperatures $T_c$ of M$_4$H (M=Ca, Cu, Rh) have a remarkable enhancement compared with the corresponding fcc metal calcium and copper (not a superconductor) and rhodium (an ultralow critical temperature superconductor). The key superconducting parameters of binary M$_4$H (M=Ca, Cu, Rh) and ternary AHM$_3$ (A=V, Nb, M=Al, Rh, Ca) perovskite hydrides with H atom at the O$\rm _h$ site are summarized in Table~{\color{blue}\ref{tables3}}. For the sake of comparison, the superconducting parameters of the binary metal hydride $\rm {Al_4H}$ with H atom at the tetrahedral site T$_{\rm d}$ are also given.

\begin{figure}[htb]
\centering
\includegraphics[width=15cm]{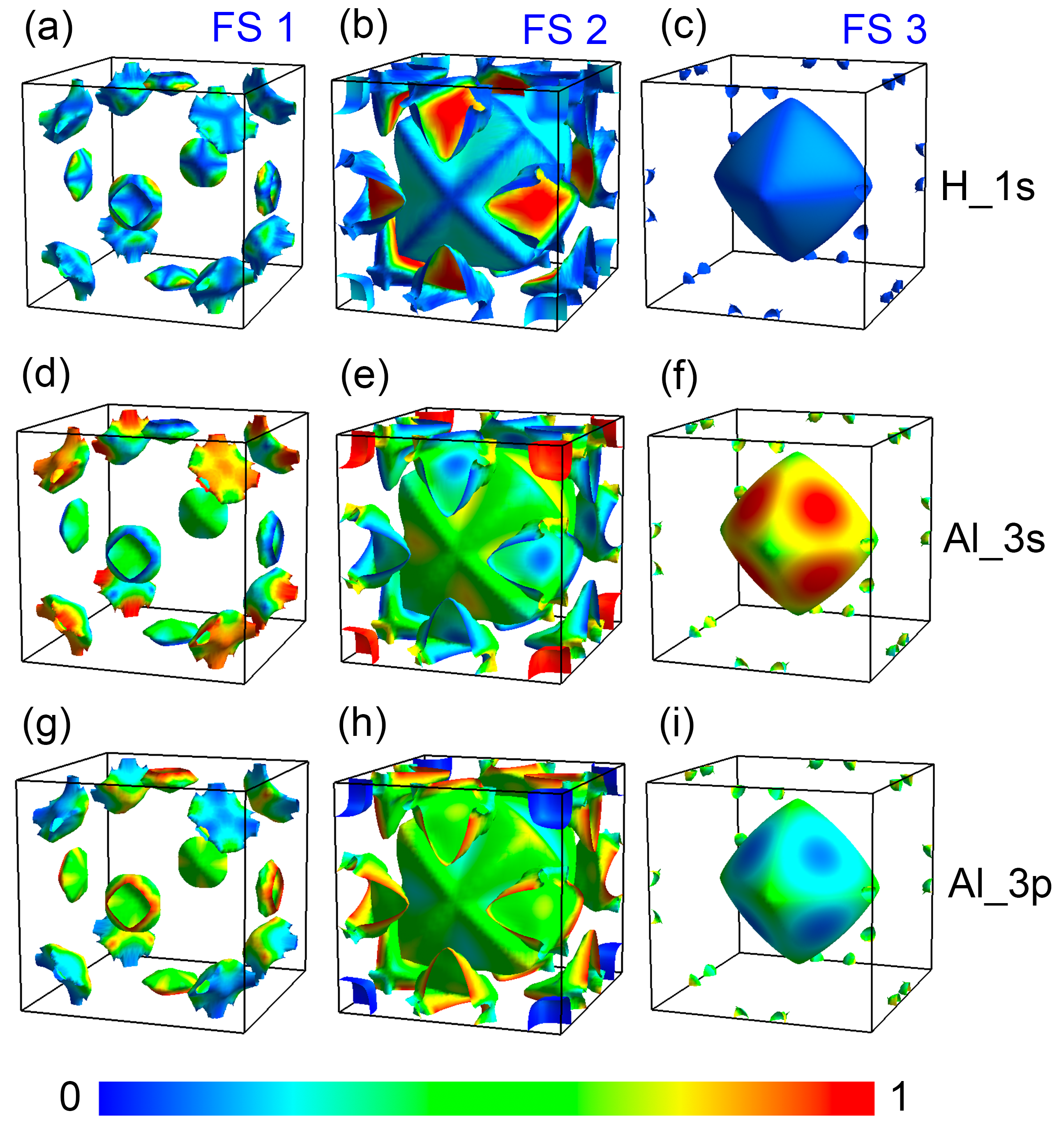}
\caption{\label{Fig_S10}Spectral weight on the Fermi surface (FS) for (a)-(c) H 1$s$ orbital, (d)-(f) Al 3$s$ orbital and (g)-(i) Al 3$p$ orbital in perovskite $\rm {Al_4H}$.}
\end{figure}

\begin{figure}[htb]
\centering
\includegraphics[width=8cm]{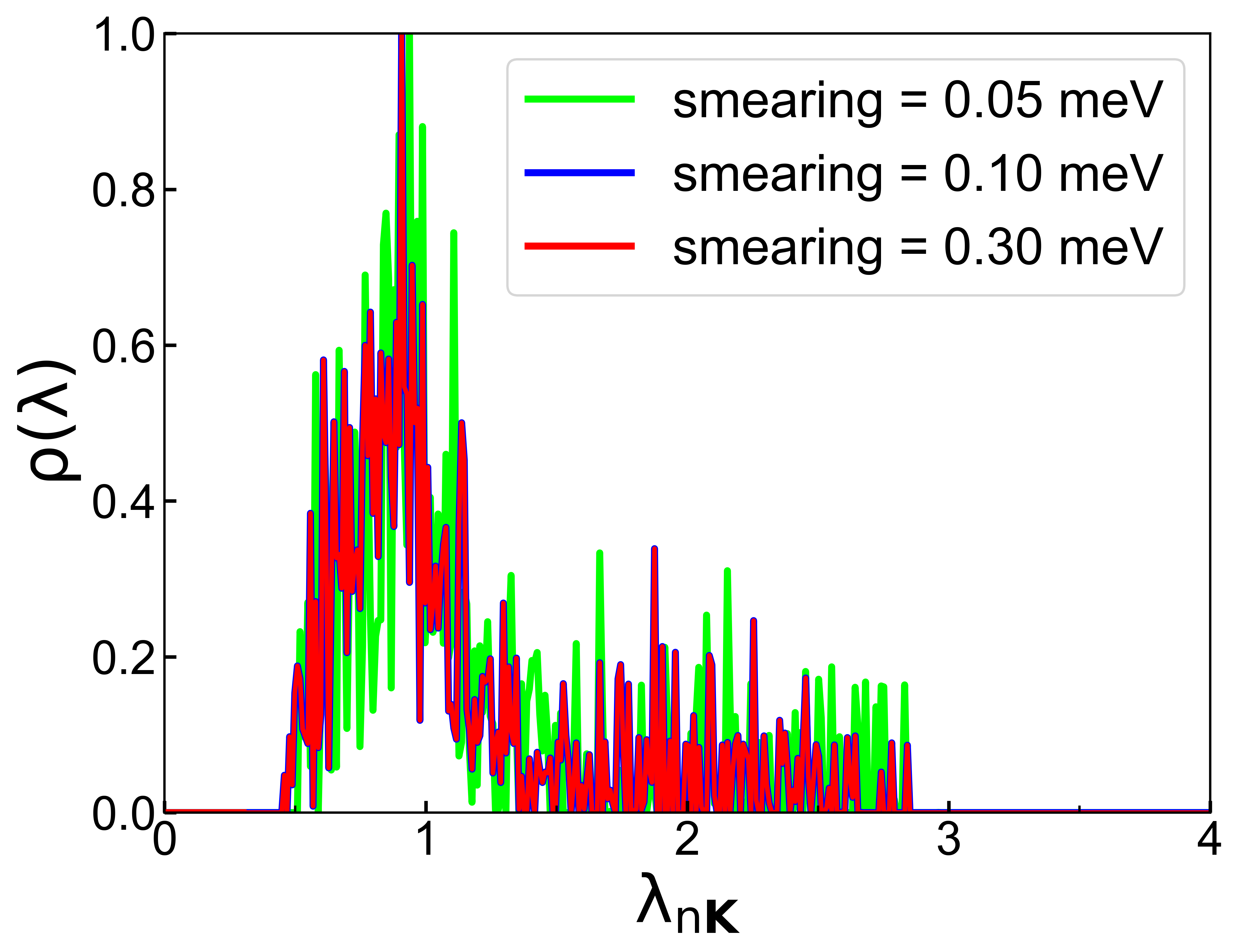}
\centering
\caption{\label{Fig_S11}~The normalized distribution of the EPC strength ${\lambda}_{n\mathbf{k}}$ in perovskite $\rm {Al_4H}$, where the green, blue and red solid lines represent the phonon smearing parameters of 0.05, 0.10 and 0.30 meV, respectively. It clearly shows that the distribution derived from the phonon smearing parameter 0.10 meV is almost identical to that of phonon smearing parameter 0.30 meV. Therefore, the phonon smearing parameter 0.30 meV achieves a sufficient convergence in our calculations.}
\end{figure}

\begin{figure}[htb]
\centering
\includegraphics[width=8cm]{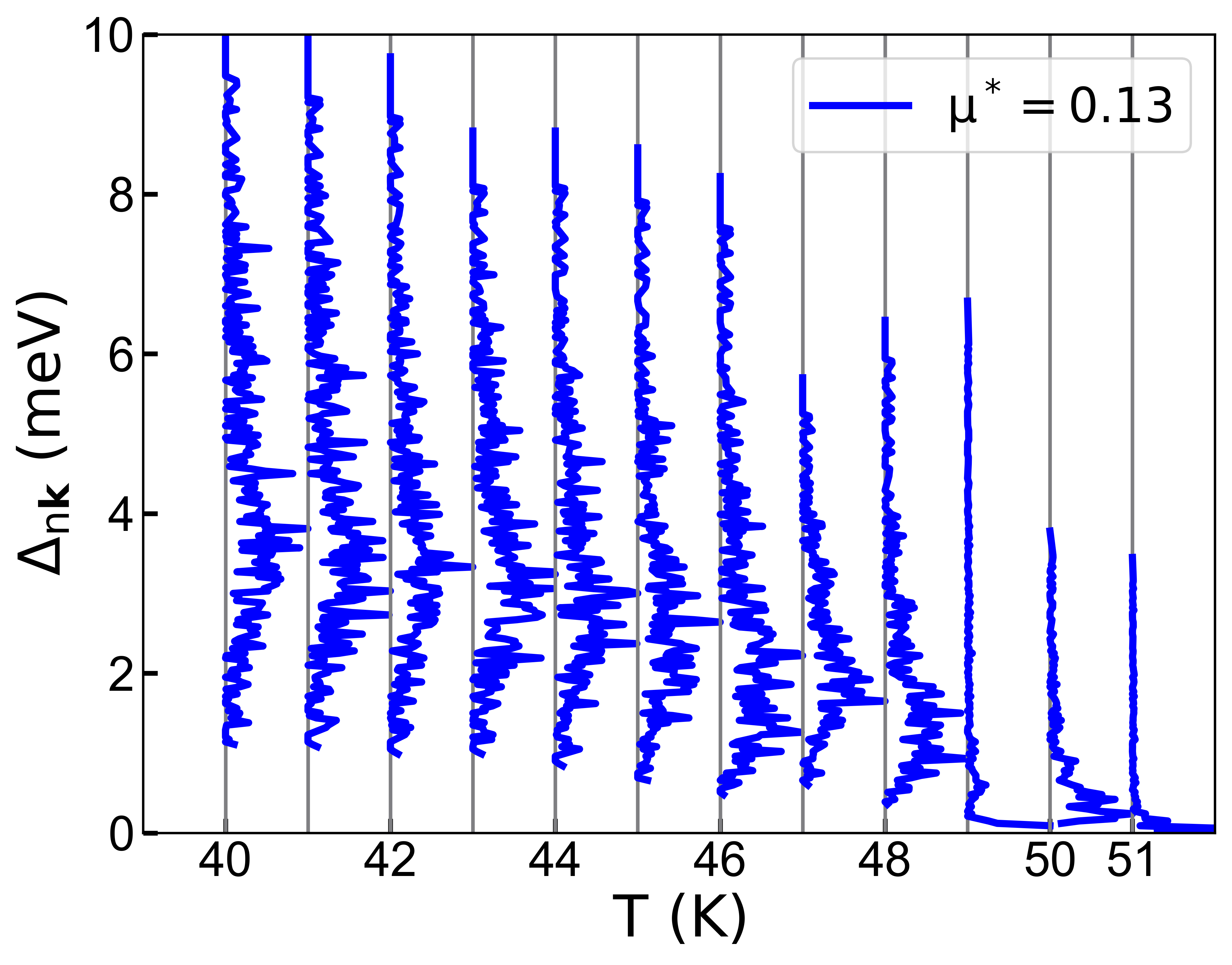}
\caption{\label{Fig_S12}~Temperature dependence of the anisotropic superconducting gap $\rm {{\Delta}_{n\mathbf{k}}}$ in perovskite $\rm {Al_4H}$, where the semiempirical Coulomb repulsion pseudopotential ${\mu}^*$ is set to 0.13. The superconducting gap $\rm {{\Delta}_{n\mathbf{k}}}$ decreases monotonously when the temperature rises and finally vanishes at the critical temperature $T_c$=51 K.}
\end{figure}

\begin{table*}[h]
\caption{\label{tables3}~Comparison of the calculated EPC strength $\rm {\lambda}$, the logarithmic average phonon frequency $\rm {{\omega}_{log}}$ and the superconducting critical temperature $T_c$ derived from  the Allen-Dynes modified McMillian formula $T_{\mathrm{c}}$=$\frac{\omega_{\log}}{1.2} \exp \left[-\frac{1.04(1+\lambda)}{\lambda-\mu^{*}(1+0.62 \lambda)}\right]$ in several  few-hydrogen metal-bonded binary perovskites M$_4$H (M=Ca, Cu, Rh) and metal hydride $\rm {Al_4H}$ with H atom at the T$_{\rm d}$ site and ternary perovskites AHM$_3$ (A=V, Nb, M=Al, Rh, Ca) with H atom at the O$_{\rm h}$ site.}
\begin{ruledtabular}
\begin{tabular}{ccccc}
\multicolumn{1}{c}{\textrm{Material}}&\textrm{$\rm {\lambda}$}&\textrm{$\rm {{\omega}_{log}}$ (K)}&\textrm{${\mu}^*$}
&\textrm{$T_c$} (K)\\
\colrule
{$\rm {Ca_4H}$ (H at O$_{\rm h}$)} & 0.30 & 223.47 & 0.10 & 0.12 \\
{$\rm {Cu_4H}$ (H at O$_{\rm h}$)} & 0.35 & 462.27 & 0.10 & 0.79 \\
{$\rm {Rh_4H}$ (H at O$_{\rm h}$)} & 0.40 & 274.08 & 0.10 & 1.17 \\
{$\rm {Al_4H}$ (H at T$_{\rm d}$)} & 0.73 & 428.48 & 0.10 & 16.23 \\
{$\rm {VHAl_3}$ (H at O$_{\rm h}$)} & 0.71 & 219.08 & 0.10 & 7.84 \\
 {$\rm {VHRh_3}$ (H at O$_{\rm h}$)} & 0.43 & 229.69 & 0.10 & 1.51 \\
 {$\rm {NbHCa_3}$ (H at O$_{\rm h}$)} & 0.28 & 165.27 & 0.10 & 0.04 \\
 {$\rm {NbHRh_3}$ (H at O$_{\rm h}$)} & 0.28 & 259.81 & 0.10 & 0.07 \\
\end{tabular}
\end{ruledtabular}
\end{table*}

\clearpage
\section{\label{sec:level6}VI. Specific heat and critical magnetic field}

\begin{figure}[htp]
\centering
\includegraphics[width=10cm]{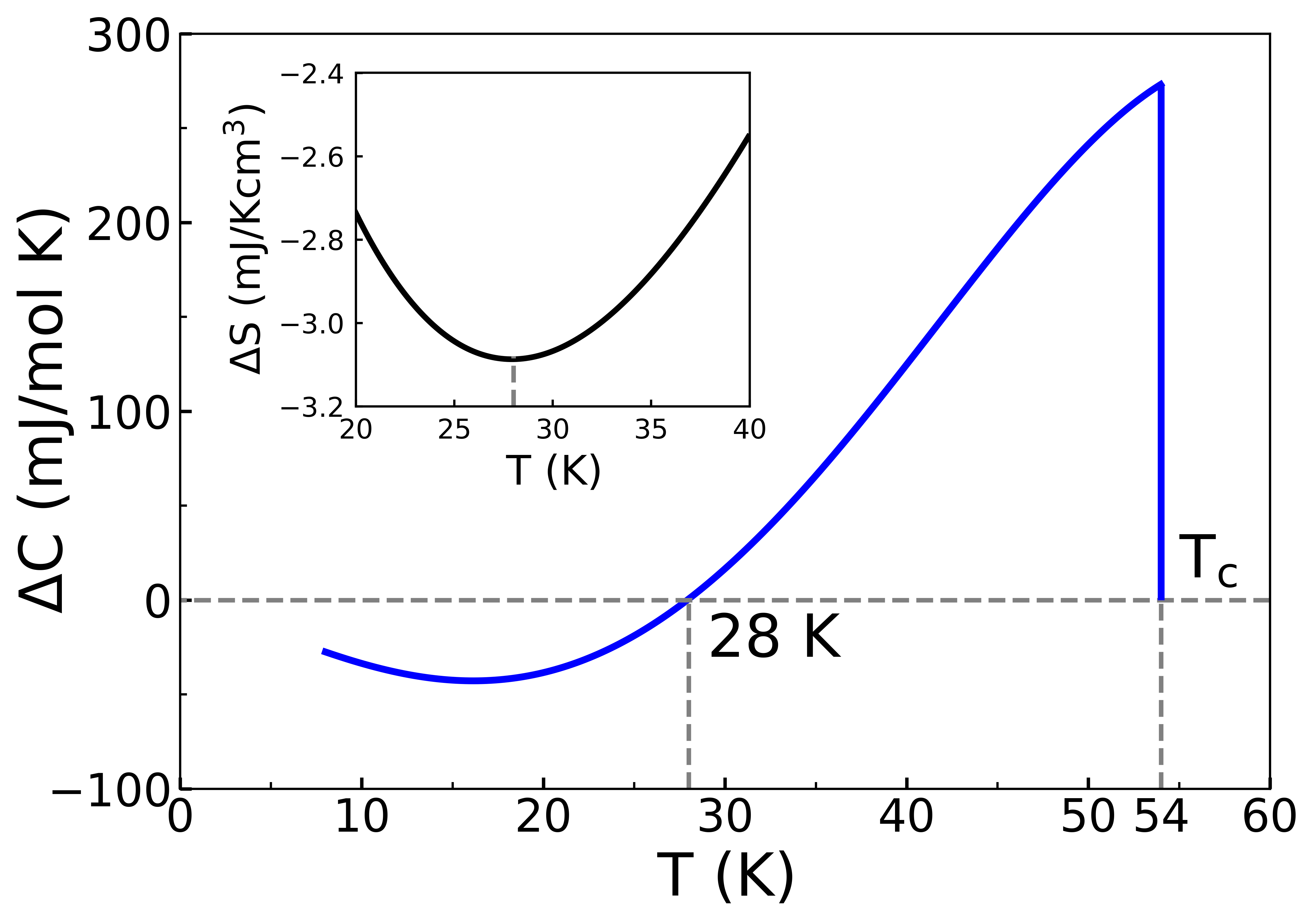}
\caption{\label{Fig_S13}Calculated specific-heat difference $C_s$-$C_n$ between the superconducting and normal states as a function of temperature by numerically computing the second-order derivative of the free energy in Al-based perovskite $\rm {Al_4H}$. The inset is its entropy difference.}
\end{figure}

The specific heat, shown in Fig.~{\color{blue}\ref{Fig_S13}}, is calculated from the second-order derivative of the free energy~(see Refs.~{\color{blue}\cite{John1964, Carbotte1990, Choi2003, Margine2013}} for calculation details), which exhibits an abrupt jump at the second-order phase transition of $T_c$=54 K, demonstrating the superconductivity of Al-based perovskite $\rm {Al_4H}$. According to the Rutgers formula, the large specific-heat jump clearly shows that $T_c$ is high and $d{\rm {H_c(T)}}/d{\rm T}$ is large, where H$\rm _c$(T) is the critical magnetic field. The inset of Fig.~{\color{blue}\ref{Fig_S13}} is the entropy difference between the superconducting and normal states. The minimum entropy difference, i.e., the zero-point of the specific-heat difference, is estimated at T=28 K, close to $T_c/\sqrt{3}$$\approx$31.18 K, which means that the formula ${\rm H}\rm_c{\rm (T)}$=${\rm H}\rm_c{\rm (0)[1-(T/T}\rm_c{\rm )^2]}$ is still applicable approximately in $\rm {Al_4H}$. Based on Eq.~{\color{blue}(\ref{equ_S6})}, we estimate the upper critical magnetic field in $\rm {Al_4H}$ is 6419 Oe, which is much larger than aluminum (100 Oe)~{\color{blue}\cite{Cochran1958}} and niobium (2038 Oe)~{\color{blue}\cite{Corsan1969}}.

\end{widetext}

\end{document}